\documentclass[preprintnumbers,amsmath,amssymb,floatfix,10pt,prd,onecolumn,superscriptaddress,nofootinbib]{revtex4}
\usepackage{epsfig,color}
\usepackage{epstopdf}
\usepackage{amssymb}
\usepackage{verbatim}
\usepackage{multirow}
\usepackage{float}
\usepackage{graphicx}
\usepackage[utf8]{inputenc}
\begin{document}
\title{Bouncing behaviours in four dimensional Einstein Gauss-Bonnet gravity with Cosmography and Observational constraints}
\author{M. Zubair}
\email{mzubairkk@gmail.com; drmzubair@cuilahore.edu.pk}\affiliation{Department
of Mathematics, COMSATS University Islamabad, Lahore Campus, Lahore,
Pakistan}
\author{Mushayydha Farooq}
\email{Mushayydha_90@hotmail.com}\affiliation{Department
of Mathematics, COMSATS University Islamabad, Lahore Campus, Lahore,
Pakistan}
\begin{abstract}

This manuscript is based on an investigation of bouncing cosmology in a 4D Einstein Gauss-Bonnet gravity. Various bouncing models such as symmetric bounce, matter bounce, super bounce, and oscillatory bounce have been examined. Expressions for energy density, pressure, equation of state parameter have been derived in the most general manner and then reduced to 4D Einstein Gauss-Bonnet gravity for isotropic, homogenous, FLRW cosmos. Physical interpretation of Hubble and deceleration parameters has also been discussed and plotted for each model from non-vanishing scale factors. Non-singular bouncing models indulge in accelerating late-time cosmic acceleration phenomenon. It has been analysed that the Gauss-Bonnet coupling parameter has a lesser contribution to the dynamics of modified gravity while the bouncing parameter has noticeable effects. We have examined various energy conditions and witnessed the violation of strong and null energy conditions in bouncing models. Analytical expressions for jerk and snap parameters have also been calculated in terms of cosmic time and redshift. We have explored bouncing models through specific cosmographic tests to check their validity. Also, through stability analysis, matter bounce becomes the most stable model by increasing the value of the bouncing parameter. To find best-fit values, bouncing models have been constrained with Hubble data set and $\Lambda$CDM. We have calculated the values of parameters by applying the least-square fitting method. To make this analysis quantified, we have employed reduced chi-squared method on $H(z)$ data sets for each model.

\end{abstract}

\date{\today}

\maketitle

\section{introduction}
Diverse experiments and various observations have been scrutinized in order to understand the theory of general relativity (GR) in strong and weak gravitational fields, and all are according to observational data sets \cite{1}. Infact, this theory also envisions us about space-time singularities under natural constraints
\cite{2}. This loophole leads us to the fact that still, we need more authentic theories that completely describe space-time, its composition,
gravity, and its whereabouts \cite{3}.
Beyond GR significant number of theories have been proposed regarding gravitation and cosmology. Including superstring or M-Theory, drawn up in higher dimensional space-time is the most favourable concept. Two particular parameters have to be introduced to make the system in the superstring theory. One parameter is the string coupling parameter which is $g^{2}_{s}=e^{\phi}$ where $\phi$ is dilation field, the second one is inverse string tension $\gamma^{'}$. When the value of  $\gamma^{'}$ is small (tension is high) in comparison to the energy scale of the system, it becomes challenging to excite strings, its size becomes very small, and it is regarded as the particle of zero-order approximation. In this limiting case, GR and other light fields can be recovered. This is called $\gamma^{'}$ expansion \cite{4}. While driving $\gamma^{'}$ higher-order terms, curvature corrections terms appear. The next level involves studying the Gauss-Bonnet (GB) term, which includes the ordinary set of equations with the maximum up to second-order derivative. In spite of, $\gamma^{'}$ expansion of type IIB superstring theory includes ghost-free combinations with higher curvature combinations \cite{5,6}. Many analysis have been performed by considering highly symmetric space-time, the system becomes much more complicated than in GR \cite{7,8,9}.

The natural generalization of Einstein gravity in higher dimension is known as Lovelock theory \cite{10}. The action is a homogeneous polynomial in Riemann curvature. It has an incredible quality that inspite of action being polynomial in Riemann curvature, still equation of motion remains second order because higher-order terms in action do not contribute to the equation where $D>2N$ here D is dimension and $N$ is the degree of curvature polynomial in action. We have experienced that Einstein action consists of many complex scalar terms made up of different combinations of matter fields and geometrical functions such as Einstein tensor, Ricci tensor, Riemann tensor. One of them is called the  GB term. The GB invariant is second-order while GR is first-order Lovelock. Therefore, we can write that Lovelock is a higher-dimensional generalization of GR. Hence, instead of using the GB term in pure form, which is a total derivative, the modified GB term, coupled with other fields, has been used.

To understand the dynamics of GB invariant, there are two useful scenarios, one of them is to couple GB invariant with scalar field while other one involves generic function of GB term. On the based of second scenario, GB modified gravity is another theory which has acquired acceptance in the last few years also known as $f(\mathcal{G})$ modified gravity \cite{11*,12*}. Through this theory we can study early as well as late times cosmological evolution by avoid ghost contributions \cite{13*}. The reconstruction and stability of $f(\mathcal{G})$ modified gravity has been discussed by \cite{14*} and they have investigated the inflationary survey by using different models. Energy conditions with different models of $f(G)$ modified gravity have also been evaluated by \cite{15*}, they have used recent updated values of the Hubble, deceleration, jerk and snap parameters to find out viability of these models. The generalization of $f(\mathcal{G})$ modified gravity has also been purposed by scientists which is called as $f(\mathcal{G},T)$ modified gravity \cite{16*}. There is another modified theory which is obtained by modifying Einstein Hilbert action (replacing $R$ by $f(R,\mathcal{G})$) \cite{17*,18*,19*} and  $f(\mathcal{G})$ gravity is its simplest form of $f(R,\mathcal{G})$. According to first scenario and recent observations, a new theory has came up with GB term \cite{20}, while GB coupling constant is scaled as $\gamma\rightarrow\dfrac{\gamma}{(D-4)}$. By substituting $D=4$ in the whole equation, it reduces into 4D. In this way GB term  contributes towards the gravitational dynamics and this idea is known as 4D Einstein Gauss-Bonnet (EGB) theory. This theory would bypass the results of Lovelock's theorem and keep away form the Ostrogradsky instability \cite{21}. Black hole solutions have been investigated in 4D EGB gravity under various circumstances including a vaidya like radiating, coupled to magnetic charge, nonlinear electrodynamics \cite{22, 23, 24, 25, 26, 27, 28, 29}. Moreover investigation have also been made to understand quasi-normal modes, deflection of light and shadow cast by black holes \cite{30,31,32}. In 4D EGB exact spherically symmetric wormhole solution for isotropic and anisotropic matter have been evaluated by considering radial space function and power law density profile \cite{34}. Moreover, the possible reconstruction of strange stares have been investigated in quark matter phases with in the background of 4D EGB \cite{36} and find out that GB term shows nontrivial contribution in the dynamics of gravitation. Electrically charged Quarks stares with static spherically symmetric spacetime have also been explored, impact of GB coupling constant on mass-radius have calculated \cite{37}. The cosmological implications of constrained EGB gravity have been evaluated in 4D and author have concluded that matter density falls more frequently at larger values of redshifts \cite{39}.


One of the most crucial cosmological problems is a cosmological singularity which is somehow resolved by the introduction of bouncing cosmology \cite{40, 41}. It has been found that bouncing cosmologies are a substitute for standard inflationary theories \cite {44, 45}. Many efforts have been put to study the bouncing cosmology in the framework of different modified theories. An investigation has been made based on bouncing cosmology in Teleprallel Gravity (TG). In this scenario, through detailed analysis,it has been observed that bouncing cosmologies turn up as a natural outcome in various early universe frameworks \cite{46,47,48,49}. In $f(T)$ gravity, possibilities of matter bounce cosmology has also been studied. Many attempts have been made on effective field theory of loop quantum gravity in TG, which gives reliable results and aligned with BICEP2 data and Planck’s experimental data \cite{50, 51}. Besides this, number of researchers investigated  bouncing cosmology with GB invariant theories and leads us to the several models in which bouncing cosmologies can result in early universe scenarios \cite{52,53,54,55}. Non-singular bouncing cosmology has been presented by using scalar matter with non-standard kinetic term \cite{56}. Authors used standard matching conditions and conclude that spectral index remains the same during the bounce. The review of success of Inflationary Cosmology, String Gas Cosmology and how cosmological fluctuations corresponds to current data generated by these two cosmological models have been discussed by \cite{57}. At classical quantum level NEC is violated and cyclic bouncing cosmology scenarios are not possible while all others remain valid \cite{58}. Recently, non-singular bouncing cosmology and scale-invariant power spectrum have been discussed in \cite{59,60}, where authors utilized single scalar field coupled with gravity in the background DHOST theories. Moreover, the relationship of bouncing models with necessary parameters in DHOST cosmology has also been investigated in \cite{61}. The bouncing solutions have been explored by considering logarithmic trace term and linear trace term in $f(G,T)$ modified gravity and it was concluded that NEC and SEC are violated \cite{62}. Authors being motivated from \cite{52}, bouncing solutions have also been studied in $f(T,B)$ modified theory by considering different types of gravitational Lagrangians and bouncing models \cite{63}. Similarly, the cosmological matter bounce model has also been discussed in the framework of symmetric teleparallel gravity $f(Q)$ with two gravitational Lagrangians and they have examined their stability and energy conditions \cite{64}. In addition to this, matter bounce model has also been used in reconstruction of $f(R,T)$ modified gravity \cite{65}. The exponential and power law bouncing models have been used to reconstruct $f(R)$ and $f(G)$ modified gravity, further second order polynomial is constructed to check stability \cite{66, 67}.

The present study attempts to explore the bouncing models in the framework of 4D EGB with a flat, isotropic FRW universe. Sections of the present analysis are organized as follows. Section \textbf{II}: It consists of the basic formulation of higher and 4D EGB gravity. Section \textbf{III}: In this section, four bouncing models, namely symmetric bounce, matter bounce, super bounce and oscillatory cosmology have been studied in 4D EGB gravity. In section \textbf{IV} and \textbf{V}, we have discussed the energy conditions the cosmography of bouncing models in terms of cosmic time and red shift. Moreover, stability of bouncing models is evaluated in section \textbf{VI}. Section \textbf{VII}: In this section, we have fitted the bouncing models with observational data sets.
Last section \textbf{VIII} concludes our findings.

\section{Basic Formulation of Friedmann Equation in 4D Einstein Gauss-Bonnet Gravity}

In D-dimensional space time EGB gravity can be derived by following action \cite{20}
\begin{equation}\label{2}
\mathbb{I}_\mathcal{G}=\int d^{D}x\sqrt{-g}\bigg[\dfrac{M_{p}^{2}R}{2}+\dfrac{\gamma\mathcal{G}}{D-4}\bigg]+\mathcal{S}_{matter},
\end{equation}
Here $g$ is determinant of $g_{\alpha\beta}$, $R$ is Ricci scalar which provides information about GR part of action. $M_{p}$ is reduced Plank mass in terms of gravitational constant i.e. $M_{p}=\sqrt{\dfrac{1}{8 \pi G}}$ where $(M_{p}=2.436\times10^{18} GeV)$. Second term in action indicate GB action, $\mathcal{G}$ is GB constant and $\gamma$ is GB coupling constant. Third term contains baryonic and dark matter components.
GB term is defined as \cite{11*}
\begin{equation}\label{3}
\mathcal{G}=R^{\alpha\beta\mu\nu}R_{\alpha\beta\mu\nu}-4R^{\alpha\beta}R_{\alpha\beta}+R^{2},
\end{equation}
where $R_{\alpha\beta}$ is Ricci tensor and $R_{\alpha\beta\mu\nu}$ is Riemann tensor.
By varying equation (\ref{2}) with respect to metric tensor $g_{\alpha\beta}$ required equation is
\begin{equation}\label{4}
G_{\alpha\beta}=\dfrac{1}{M_{p}^{2}}\bigg(\dfrac{\gamma\mathcal{G}}{D-4}H_{\alpha\beta}+T_{\alpha\beta}\bigg),
\end{equation}
where, $G_{\alpha\beta}$ is Einstein tensor, $H_{\alpha\beta}$ is Lancoz tensor, where expression for energy momentum tensor for $\mathcal{S}_{matter}$ is defined as
\begin{equation}\label{5}
T_{\alpha\beta}=-\dfrac{2}{\sqrt{-g}}\dfrac{\delta(\sqrt{-g}\mathcal{S}_{matter})}{\delta g^{\alpha\beta}},
\end{equation}
Expression for Lancoz tensor is
\begin{equation}\label{6}
H_{\alpha\beta}=2\bigg(RR_{\alpha\beta}-2R_{\alpha\mu}R^{\mu}_{\beta}-2R_{\alpha\mu\beta\nu}R^{\alpha\beta}-R_{\alpha\mu\nu\delta}R^{\mu\nu\delta}_{\beta}\bigg)-\dfrac{1}{2}\mathcal{G}g_{\alpha\beta},
\end{equation}
In D-dimensional, flat FLRW space time is defined as
\begin{equation}\label{7}
ds^{2}=a^{2}(t)dx_{1}^{2}+a^{2}(t)dx_{2}^{2}+a^{2}(t)dx_{3}^{2}+a^{2}(t)dx_{4}^{2}+......-dt^{2}.
\end{equation}
where $a(t)$ is cosmic scale factor which reveals about expension of universe. It is dimensionaless and key parameter for FLRW space time.
GB scalar term for FLRW is derived as
\begin{equation}\label{8}
\mathcal{G}=(D-1)(D-2)(D-3)4H^{2}(\dot{H}+H^{2})+(D-1)(D-2)(D-3)(D-4)H^{4}.
\end{equation}
By variation of action (\ref{2}) with respect to $g_{\alpha\beta}$, following non-zero components are obtained,
\begin{eqnarray}
\dfrac{(D-1)(D-2)H^{2}}{2}&=&-\dfrac{\gamma H^{4}}{M_{p}^{2}}(D-1)(D-2)(D-3)(D-4)+\dfrac{\rho}{M_{p}^{2}},\\\nonumber
-\dfrac{(D-2)(D-3)H^{2}}{2}-\dfrac{(D-2)\ddot{a}}{a}&=&\dfrac{p}{M_{p}^{2}}+\dfrac{\gamma}{M_{p}^{2}}\bigg[(D-4)(D-3)(D-2)(D-5)H^{4}+\\\label{10}4(D-4)(D-3)(D-2)4\dfrac{\ddot{a}}{a}H^{2}\bigg].
\end{eqnarray}
Here, Hubble parameter is expressed as $H=\dfrac{\dot{a}}{a}$, $\rho$ is energy sensity and $p$ is pressure. Eq. (\ref{10}) can be re-written as
\begin{equation}\label{11}
\dot{H}=-\dfrac{\rho+p}{4\gamma(D-2)(D-3)(D-4)H^{2}+(D-4)M_{p}^{2}}.
\end{equation}
The field equations of D-dimensional EGB gravity in FLRW space time for energy density $\rho$, $p$ and equation of state parameter (EoS) $w=\dfrac{p}{\rho}$ are expressed as
\begin{eqnarray}
\rho& = &\gamma(D-1)(D-2)(D-3)(D-4)H^{4}+\dfrac{(D-1)(D-2)M_{p}^{2}H^{2}}{2},\\\label{12}
p&=&-\gamma(D-1)(D-2)(D-3)(D-4)H^{4}-\dfrac{(D-1)(D-2)M_{p}^{2}H^{2}}{2}
-\dot{H}(D-2)\bigg(4\gamma(D-3)(D-4)H^{2}+M_{p}^{2}\bigg).\\\label{13}
w&=&-1-\dfrac{\dot{H}(D-2)\bigg(4\gamma(D-3)(D-4)H^{2}+M_{p}^{2}\bigg)}{\gamma(D-1)(D-2)(D-3)(D-4)H^{4}+\dfrac{(D-1)(D-2)M_{p}^{2}H^{2}}{2}}.\label{14}
\end{eqnarray}
These equations can be reduced into 4D by putting  $\gamma(D-4)$ equal to finite non-zero value. This is only possible, by setting $D=4$ and $\gamma$ $\rightarrow$
$\dfrac{\gamma}{(D-4)}$. Thus Energy density, pressure and EoS parameter are reduced as follows
\begin{eqnarray}
\rho&=&3M_{p}^{2}H^{2}+6\gamma H^{4},\\
p&=&-3M_{p}^{2}H^{2}-6\gamma H^{4}-2\dot{H}\bigg(4\gamma H^{2}+M_{p}^{2}\bigg),\\
w&=&-1-\dfrac{2\dot{H}\bigg(4\gamma H^{2}+M_{p}^{2}\bigg)}{3M_{p}^{2}H^{2}+6\gamma H^{4}}.
\end{eqnarray}

In 4D FLRW space, Ricci scalar $R$ and GB term $\mathcal{G}$ are defined as $R=6(\dot{H}+2H^{2})$ and $ \mathcal{G}=24H^{2}(H^{2}+\dot{H})$. We can mainly assume from here that modification in cosmic evolution depends on $\rho$, $p$ and $w$. It is obvious that the dynamical behaviour of these parameters depends on GB coupled parameter $\gamma$. For $\gamma=0$, we can get EoS parameter in GR.

\section{Bouncing behaviour in 4D Einstein Gauss-Bonnet gravity}

In the present analysis, we intend to discuss various bouncing scenarios in 4D EGB gravity. This investigation comprises of, dynamics of energy density $\rho$, pressure $p$ and EoS parameter $w$. The preceding section consists of their expression in D and 4D EGB gravity connected with the Hubble parameter. Generally, the following conditions have been satisfied by bouncing models. \\

$\bullet$ Bouncing models experiences contracting phase before, leading to non-singular bounce \emph{i.e.}, the expansion of universe $a(t)$ decreases with time as $\dot{a}(t)<0$. Therefore, Hubble parameter $H=\dfrac{\dot{a}}{a}<0$ represents contracting era of universe.

$\bullet$ The scale factor contracts to zero at bouncing point  $\dot{a}(t)=0$. Accordingly, the Hubble parameter vanishes at bouncing point $H=0$. For homogenous and flat FRW, the EoS $w$ and deceleration parameter $q$ are expressed as $w=-1-\dfrac{2\dot{H}}{3H^{2}}$ and $q=-1-\dfrac{\dot{H}}{H^{2}}$ respectively. It can be seen that, at the bounce point both expressions show singular behaviour.

$\bullet$ After the bouncing point scale factor $a(t)$ starts accelerating with increase in cosmic time $t$ this implies $\dot{a}(t)>0$ and therefore $H>0$. We can predict that close to the bouncing point acceleration should give rise to positive values of derivative of $H$ $(\dot{H}>0)$. For a bouncing scenario, the EoS parameter evolves in a phantom era.

This segment of manuscript consists of dynamics of geometric parameters such as scale factor, Hubble parameter and deceleration parameter in 4D EGB gravity. Here, the contribution of theory can be measured by GB coupled parameter $\gamma$ while bouncing effects can be measured by bouncing parameter. This section consists of four bouncing models: symmetric bounce, matter bounce, super bounce, and oscillatory bounce. Furthermore, the behaviour of energy density, pressure and EoS parameter have also been studied in terms of cosmic time.\\

\subsection{Symmetric Bounce}

The symmetric bounce can be pictured through exponential scale factor as \cite{70,71},
\begin{equation}\label{18}
a(t)=e^{\lambda t^{2}},
\end{equation}
Positive values of $\lambda$ control cosmic expansion and negative values of $\lambda$ control cosmic contraction. As the Universe is expanding day by day, there is no need to discuss the contraction phase, therefore we set $\lambda > 0$. Bouncing point appears at $t=0$. The Hubble parameter is expressed as
\begin{equation}\label{19}
H(t)=2\lambda t.
\end{equation}

\begin{figure}[th!]
\centering \epsfig{file=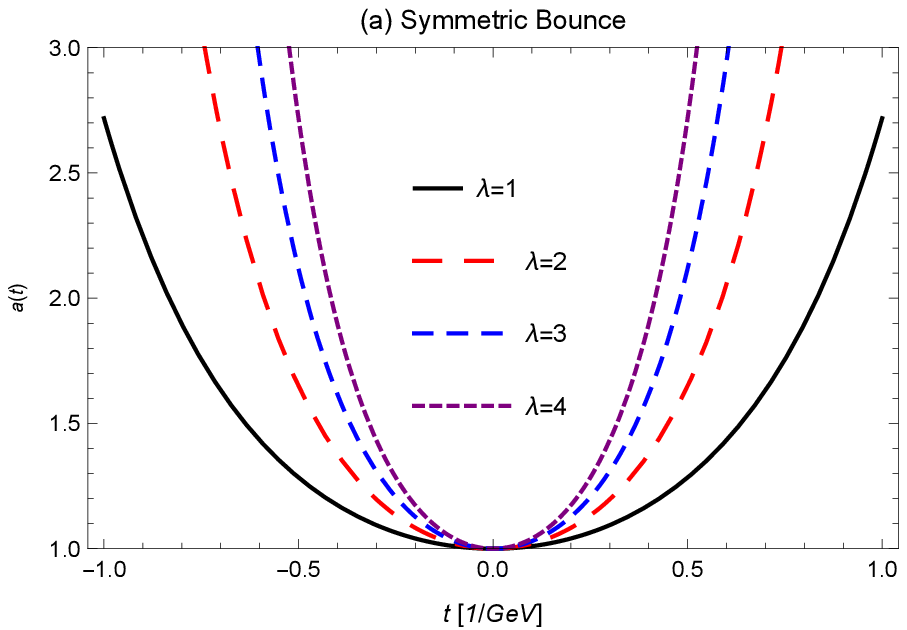, width=.35\linewidth,
height=1.51in}\epsfig{file=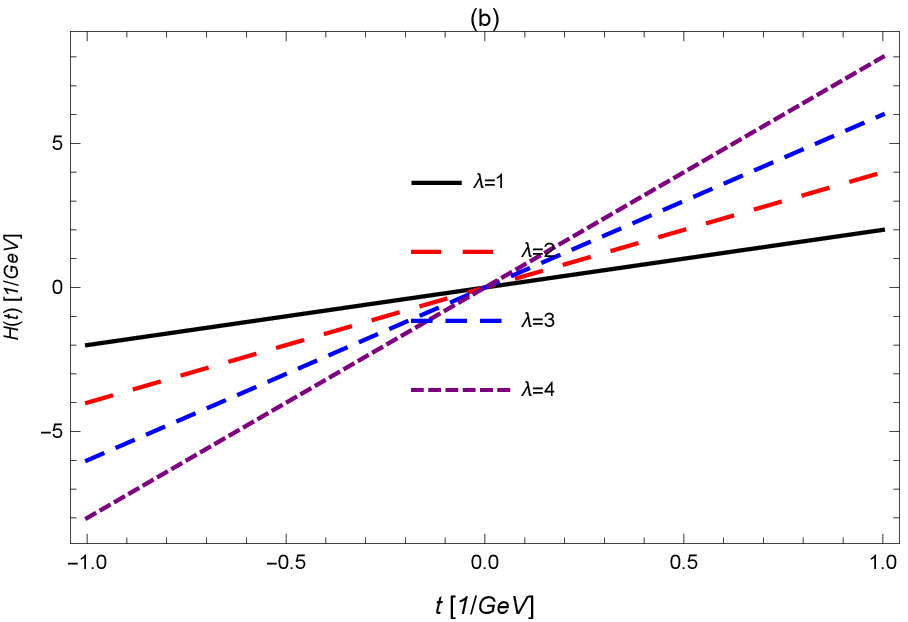, width=.35\linewidth,
height=1.51in}\epsfig{file=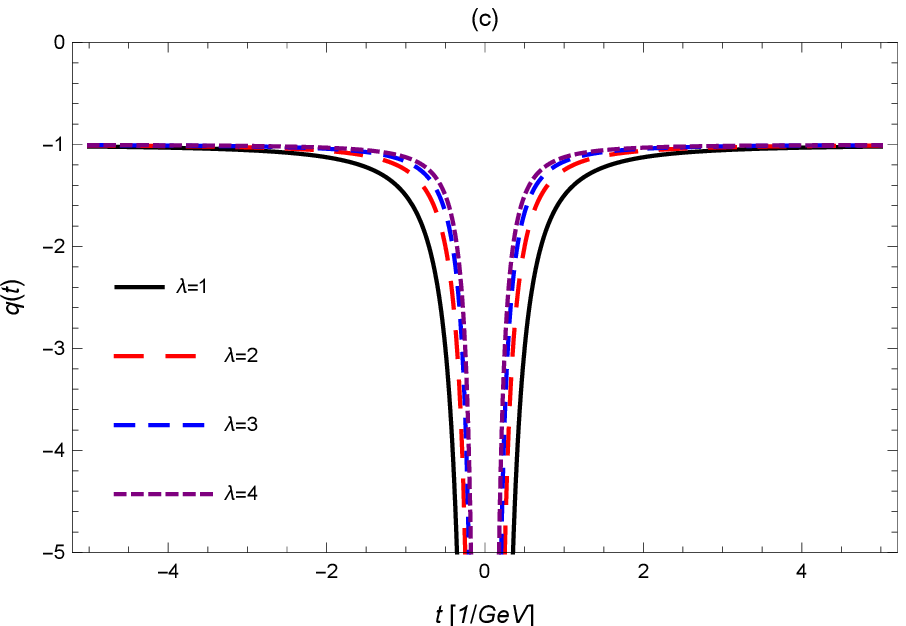, width=.35\linewidth,
height=1.51in}\caption{\label{Fig.1} Evolution of $a(t)$, $H(t)$ and $q(t)$ against $t$ for different values of $\lambda$.}
\end{figure}

Figure 1(a) shows evolutionary behaviour of scale factor for four different values of bouncing parameter $\lambda= 1,2,3,4$. It can be predicted that scale factor shows symmetric behaviour for $t<0$, $t>0$ and $t=0$ is bouncing epoch. With the increase in the bouncing parameter $\lambda$, scale factor expands. It can be observed that curvature of curves are directly proportional to parameter $\lambda$. Figure 1 (b) shows behavior of $H(t)$ for four different values of bouncing parameter $\lambda$. In this case $H(t)$ comprises of positive and negative values of cosmic time therefore, the Hubble parameter ranges from negative domain to positive domain and the bouncing parameter only accelerates the numeric values of $H(t)$.
The deceleration parameter in terms of Hubble parameter is defined as
\begin{equation}\label{20}
q+1=-\dfrac{\dot{H}}{H^{2}},
\end{equation}
For symmetric scale factor deceleration parameter is defined as
\begin{equation}\label{21}
q(t)=-1-\dfrac{1}{2 \lambda t^{2}}.
\end{equation}
The positive range of the $q(t)$ specifies the decelerated universe and negative range predicts the accelerated universe. Figure 1 (c) shows negative values of deceleration parameter $q(t)$ for all values of cosmic time $t$  therefore, symmetric bouncing model always predicts accelerated phase. It can be observed from Figure 1 (c) that $q(t)$ shows symmetric behaviour around $t=0$. For very small values (large negative) of cosmic time $t$, $q(t)$ stays at $q=-1$, its value gradually decreases close to the bouncing point. For large positive values of cosmic time $t$, $q(t)$ ranges at $q=-1$ then gradually decreases  to large negative values near the bouncing point.
For present model energy density $\rho(t)$, pressure $p(t)$ and EoS parameter $w(t)$ are expressed as
\begin{eqnarray}
\rho(t) &= &12M_{p}^{2}t^{2}\lambda^{2}+96t^{4}\gamma\lambda^{4},\\\label{22}
p(t)&=&-12 M_{p}^{2}t^{2}\lambda^{2}-96t^{4}\gamma\lambda^{4}-4\lambda\bigg(M_{p}^{2}+16t^{2}\gamma\lambda^{2}\bigg),\\\label{23}
w(t)&=&-1-\dfrac{4\lambda\bigg(M_{p}^{2}+16t^{2}\gamma\lambda^{2}\bigg)}{12M_{p}^{2}t^{2}\lambda^{2}+96t^{4}\gamma\lambda^{4}}.\label{24}
\end{eqnarray}

\begin{figure} [th!]
	\epsfig{file=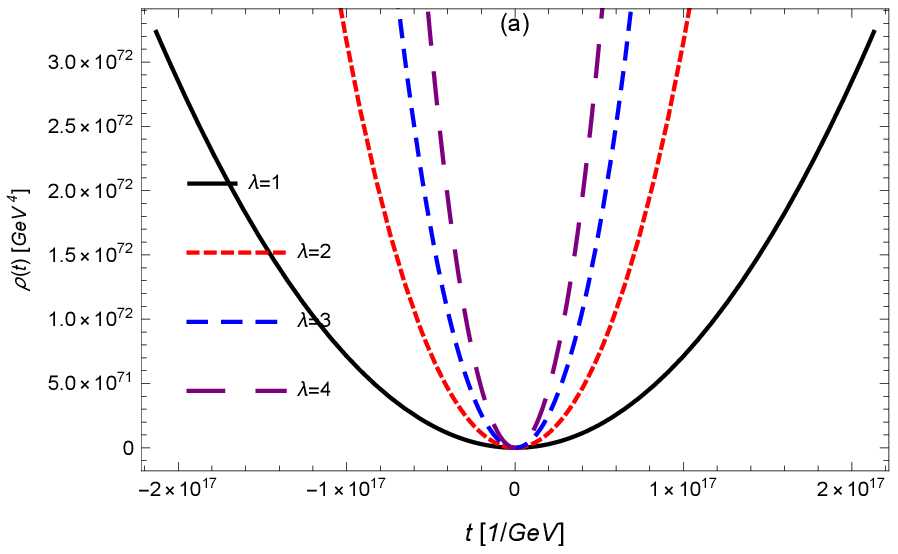,width=0.47\linewidth}\epsfig{file=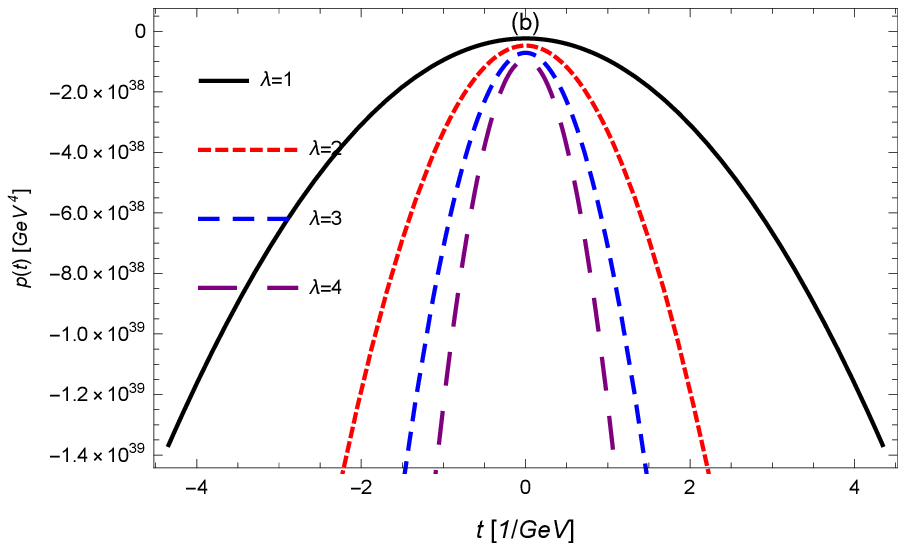,width=0.47\linewidth}
	\caption{The left plot (a) corresponds to the behaviour of $\rho(t)$, whereas right plot (b) depicts the behaviour of $p(t)$. Both graphs are plotted versus $t$ for different values of $\lambda$ and $\gamma=2$.}\label{fig2}
\end{figure}

\begin{figure} [th!]
	\epsfig{file=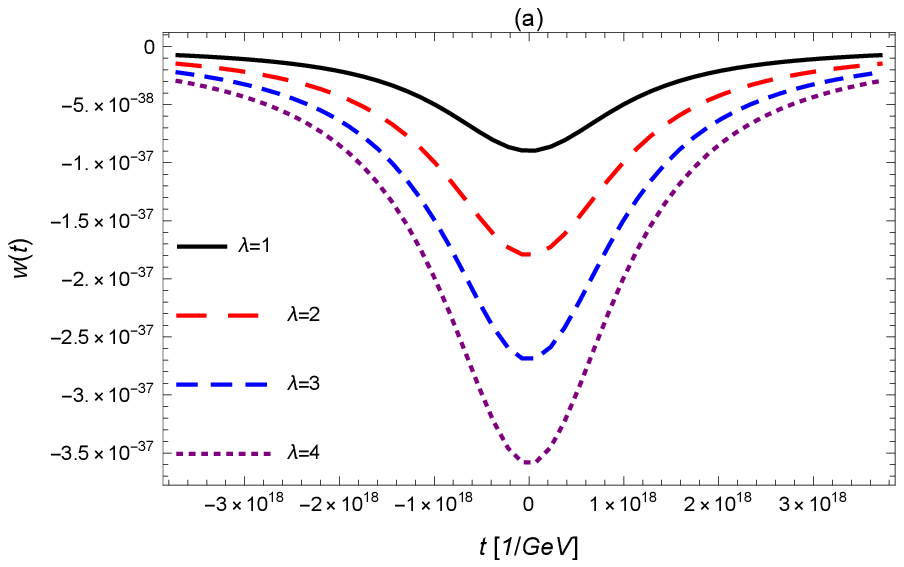,width=0.47\linewidth}\epsfig{file=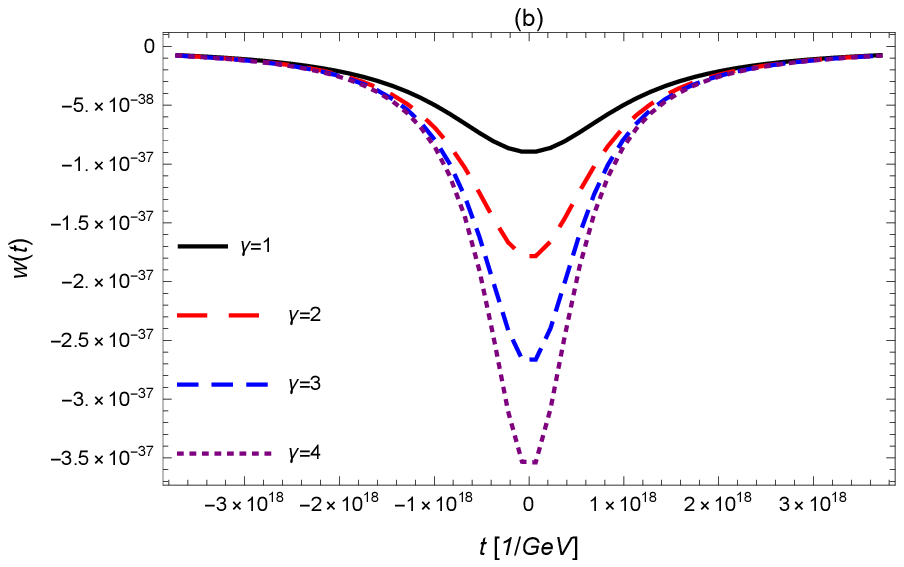,width=0.47\linewidth}
	\caption{The plot (a) corresponds to the behaviour of $w(t)$ against $t$ when $\gamma=0.1$ and $\lambda$ is varying, whereas graph (b) depicts the behaviour $w(t)$ against $t$ when $\lambda=1$ while $\gamma$ is not fixed.}\label{fig3}
\end{figure}
Expression for energy density $\rho(t)$ and pressure $p(t)$ shows its dependence on the bouncing parameter $\lambda$ and GB coupled parameter $\gamma$. It can be observed from energy density expression that it will always remain positive for both $\lambda<0$ and $\lambda>0$ as expression contains even powers. Figure 2 (a) and (b) depicts the behaviour of energy density $\rho(t)$ and pressure $p(t)$, with an increase in bouncing parameter curvature of energy density curves increases while in the case of pressure same behaviour is experienced with negative range. Evolutionary behaviour of EoS parameter againt cosmic time for different choices of $\lambda$ and $\gamma$ is displyed in Figure 3 (a) and (b). If $\gamma$ is subtituted equal to zero in equation (\ref{24}) then results reduced to EoS parameter in GR which is $w=-1-\dfrac{1}{3\lambda t^{2}}$. In symmetric bouncing cosmology, the EoS parameter evolves in the phantom region from large negative values to large positive values except at the bouncing point. $w$ gives rapid results near the bounce than far away from the bouncing point concerning time. The equation (\ref{24}) shows that the bouncing parameter $\lambda$ contributes significantly near the bounce than far away from the bounce. \textcolor{red}{ The EOS parameter shows singular behaviour for $t=0$ $\implies$   $w(t)\rightarrow\infty$. In order to restrict EOS parameter, we apply L'Hospital rule, we get $w=-\frac{16 \gamma  \lambda }{3 \left(M_{p}^2+48 \gamma  t^2\right)}$.
We have constrainted the $\lambda$, $\gamma$ and $M_{p}$ through the condition $w<-1$ $\implies$ $\dfrac{16\gamma\lambda}{3}<M_{p}^{2}$. By using this constraint, we have find the contribution of GB coupled parameter $\gamma$, by taking the fixed values of bouncing parameter $\lambda$ plotted in Figure 3 (a). It can be observed from Figure 3 (b) that there is an impressive contribution of $\gamma$. We have checked it at $\gamma= 1, 2, 3, 4$ up till large values of $\gamma$ and for the constant value of bouncing point $\lambda=1$.}

\subsection{Matter Bounce}
In this subsection, we have examined another bouncing scale factor in 4D EGB gravity which is defined as \cite {65}
\begin{equation}\label{26}
a(t)=\bigg(a_{0}+\alpha^{2}t^{2}\bigg)^{\dfrac{1}{2}},
\end{equation}
Here, $\alpha$ is a positive parameter as this study is for the accelerated expansion phase. $a_{0}$ is the radius of scale factor at the bouncing point. Special form of scale factor for matter bounce model $a(t)=\bigg(a_{0}+\alpha^{2}t^{2}\bigg)^{\dfrac{1}{3}}$ and $a(t)=\bigg(a_{0}+\alpha^{2}t^{2}\bigg)^{\dfrac{1}{4}}$ and other different powers $\large(\dfrac{2}{3},\dfrac{4}{3},\dfrac{3}{2},\dfrac{3}{4}\large)$ has also been studied by \cite{72, 73, 74, 75, 76}. These bouncing scenarios examine non-singular bounce coupled with the contracted matter-dominated state. Such type of models provides alternatives to inflation by reproducing observed spectrum of cosmological fluctuations. These models do not satisfy SEC near the bouncing epoch by instigating a new form of matter in the background of GR. One can analyze that one must study beyond GR to understand bouncing cosmology while keeping the matter content unmodified \cite{77*}. Figure 4 (a) shows the behaviour of scale factor against cosmic time for different values of bouncing parameter $\alpha$. It can be observed that the bouncing point is at $t=0$, and it is symmetric. The slope of scale factor $a(t)$ depends on bouncing parameter  $\alpha$. In short, parameter $\alpha$ is prominent factor in controlling the slope of $a(t)$.
Expression for Hubble parameter is defined as
\begin{equation}\label{27}
H(t)=\dfrac{t\alpha^{2}}{a_{0}+\alpha^{2}t^{2}}.
\end{equation}
Figure 4 (b) shows variation of Hubble parameter against cosmic time for different values of bouncing parameter. The deceleration parameter for present model is defined as
\begin{equation}\label{28}
q(t)=-\dfrac{a_{0}}{t^{2}\alpha^{2}}.
\end{equation}
Figure 4 (c) expresses the behaviour of deaccelertion parameter, its negative range shows accelerated phase of expansion.
\begin{figure}[th!]
\centering \epsfig{file=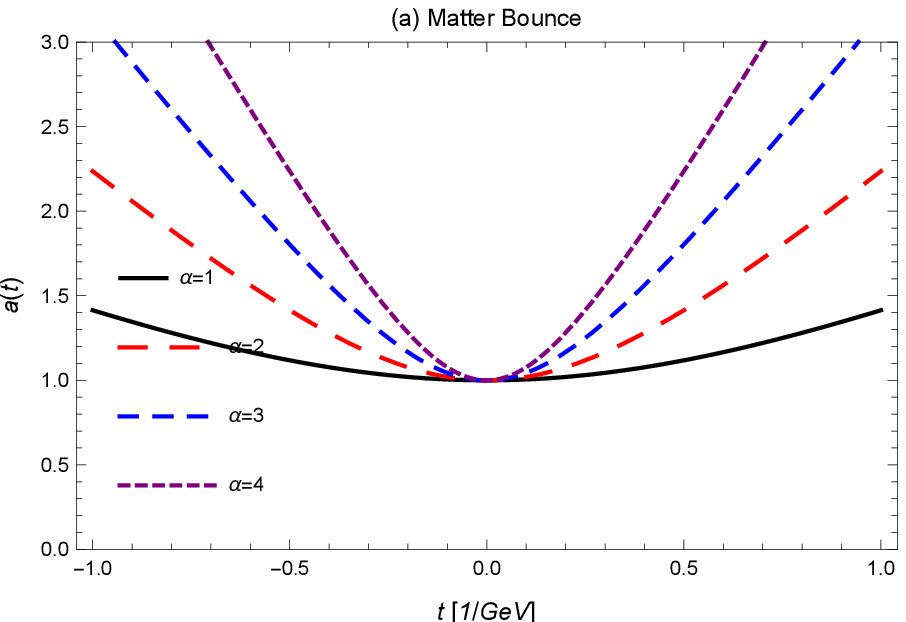, width=.35\linewidth,
height=1.51in}\epsfig{file=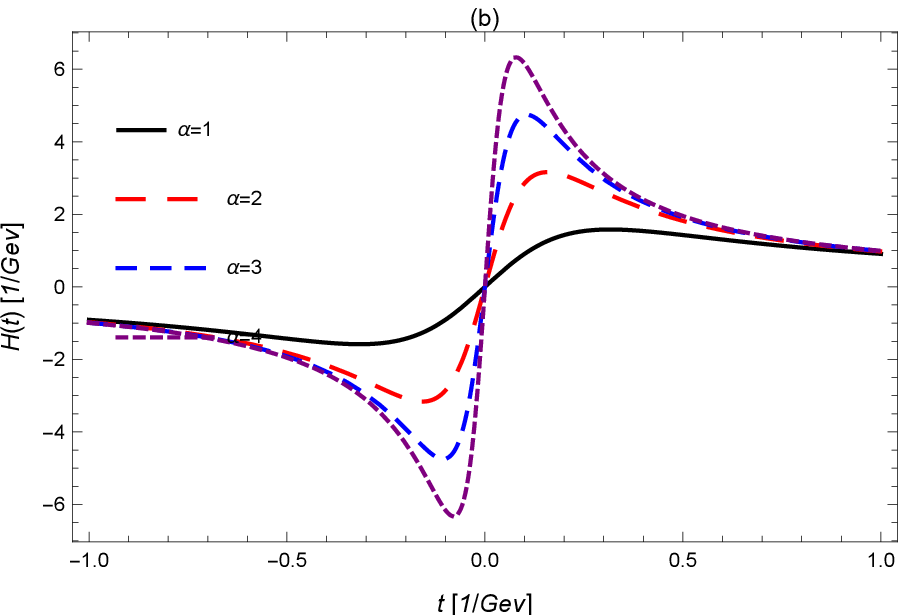, width=.35\linewidth,
height=1.51in}\epsfig{file=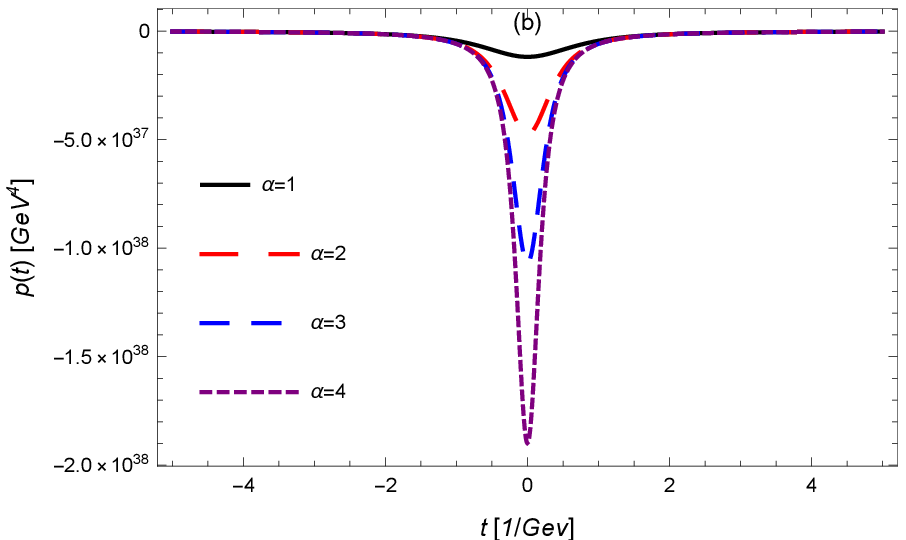, width=.35\linewidth,
height=1.51in}\caption{\label{Fig.4} Evolution of $a(t)$, $H(t)$ and $q(t)$ against $t$ for different values of $\alpha$.}
\end{figure}
Energy density $\rho(t)$, pressure $p(t)$ and EoS parameter $w(t)$ for present model are as follows
\begin{eqnarray}
\rho(t)&=&\dfrac{6t^{4}\gamma\alpha^{8}}{\bigg(t^{2}\alpha^{2}+a_{0}^{2}\bigg)^{4}}+\dfrac{3M_{p}^{2}t^{2}\alpha^{4}}{\bigg(t^{2}\alpha^{2}+a_{0}^{2}\bigg)^{2}},\\\label{28}
p(t)&=&-\dfrac{6t^{4}\gamma\alpha^{8}}{\bigg(t^{2}\alpha^{2}+a_{0}^{2}\bigg)^{4}}-\dfrac{3M_{p}^{2}t^{2}\alpha^{4}}{\bigg(t^{2}\alpha^{2}+a_{0}^{2}\bigg)^{2}}-2\bigg(M_{p}^{2}+\dfrac{4t^{2}\gamma\alpha^{4}}{(t^{2}\alpha^{2}+a_{0}^{2})^{2}}\bigg)\bigg[-\dfrac{2t^{2}\alpha^{4}}{(t^{2}\alpha^{2}+a_{0}^{2})^{2}}+\dfrac{\alpha^{2}a_{0}^{2}}{(t^{2}\alpha^{2}+a_{0}^{2})^{2}}\bigg],\\\label{29}
w(t)&=&-1-\dfrac{2\bigg(M_{p}^{2}+\dfrac{4t^{2}\gamma\alpha^{4}}{\bigg(t^{2}\alpha^{2}+a_{0}^{2}\bigg)^{2}}\bigg)\bigg[-\dfrac{2t^{2}\alpha^{4}}{(t^{2}\alpha^{2}+a_{0}^{2})^{2}}+\dfrac{\alpha^{2}a_{0}^{2}}{(t^{2}\alpha^{2}+a_{0}^{2})^{2}}\bigg]}{\dfrac{6t^{4}\gamma\alpha^{8}}{\bigg(t^{2}\alpha^{2}+a_{0}^{2}\bigg)^{4}}-\dfrac{3M_{p}^{2}t^{2}\alpha^{4}}{\bigg(t^{2}\alpha^{2}+a_{0}^{2}\bigg)^{2}}}.\label{30}
\end{eqnarray}
It is observed from the above expressions that energy density and pressure are function of bouncing parameter $\alpha$ and GB coupled parameter $\gamma$. In the present work, we have considered suitable values of these parameters and plotted them against cosmic time as depicted in Figure 5 (a) and (b).
\begin{figure} [th!]
	\epsfig{file=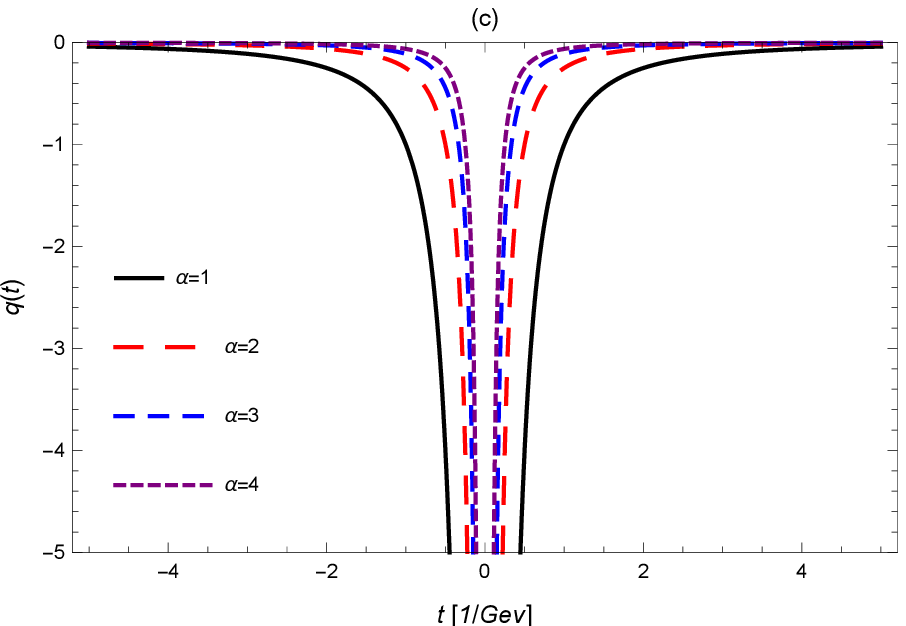,width=0.47\linewidth}\epsfig{file=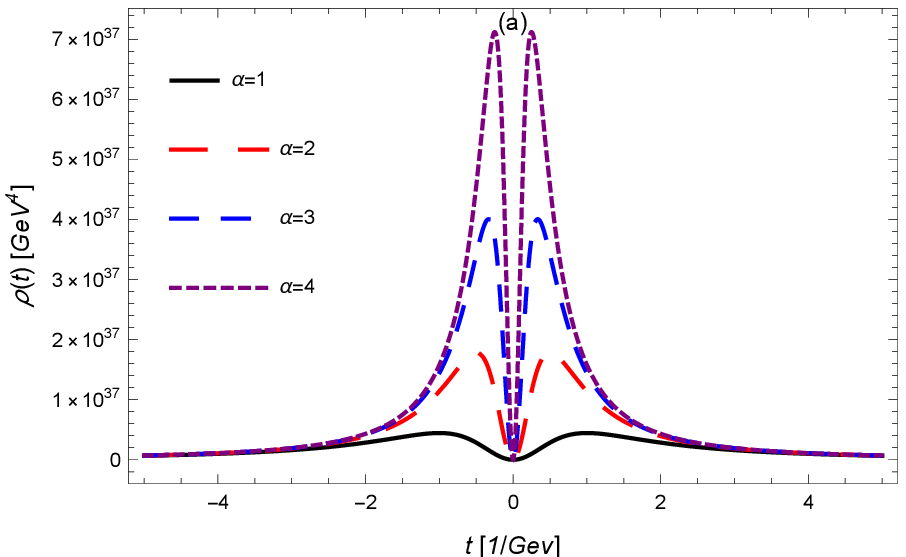,width=0.47\linewidth}
	\caption{Left plot (a) corresponds to the behaviour of $\rho(t)$, whereas right
		 graph (b) depicts the behaviour of $p(t)$. Both graphs plotted versus $t$ for different values of $\alpha$ and $\gamma=2$.}\label{fig5}
\end{figure}

\textcolor{red}{The dynamical behaviour of the EoS parameter is plotted in Figure 6 for different values of $\alpha$ and $\gamma$. The EoS parameter largely depends on the bouncing parameter $\alpha$ for non-zero constant values of $\gamma$. Results show symmetric behaviour near the bounce. Like in symmetric bounce, matter bounce has less dependence on $\alpha$ near the bouncing point. Near the bouncing point, $w$ depicts singular effects. Singular effects can be vanished by constraining EOS parameter, the condition is $\dfrac{8\gamma\alpha^{2}}{a_{0}^{2}}<M_{p}^{2}$. Figures 6 (a) and (b) show the effect of the EoS parameter $w$ on cosmic time by varying $\alpha$ and keeping $\gamma$ constant and by varying $\gamma$ keeping $\alpha$ constant, respectively. Both plots depict the phantom region as $w<-1$. It is pretty significant to understand that bouncing parameter $\alpha$ has more beneficial effects on the EoS parameter than GB coupled parameter $\gamma$.}
\begin{figure} [th!]
	\epsfig{file=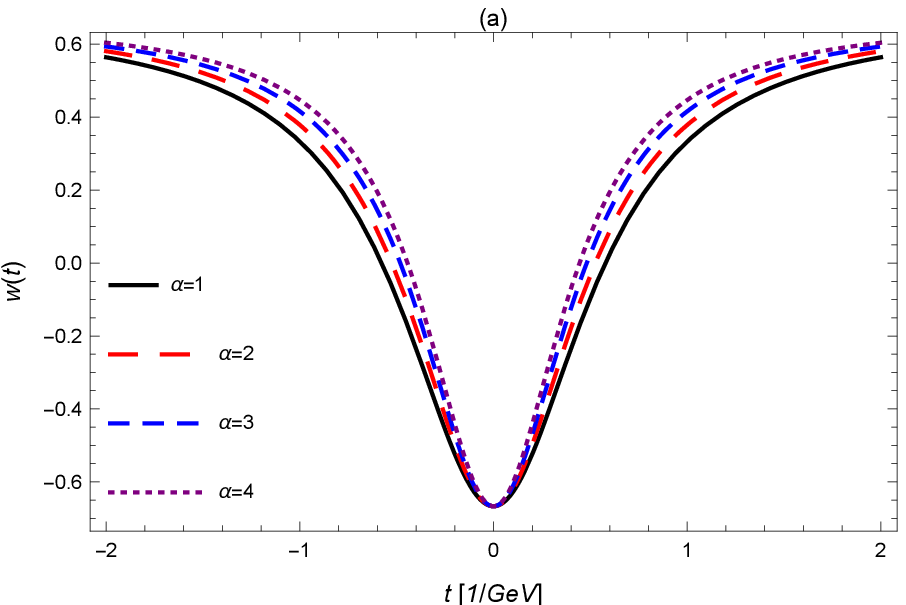,width=0.47\linewidth}\epsfig{file=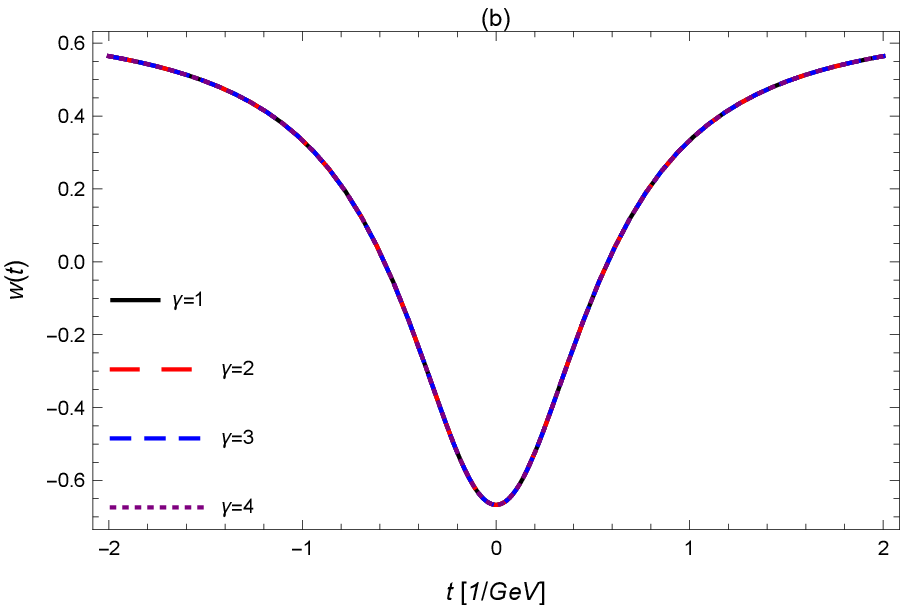,width=0.47\linewidth}
	\caption{The left plot (a) corresponds to the behaviour of $w(t)$ against $t$ when $\gamma=1$ and $\alpha$ is varying, whereas graph (b) depicts the $w(t)$ against $t$ when $\alpha=1$ is fixed while $\gamma$ is varying.}\label{fig6}
\end{figure}

\subsection{Super Bounce}
The super bounce by power law scale factor is expressed as \cite{63}
\begin{equation}\label{33}
a(t)=a_{0}+\beta t^{2n}.
\end{equation}
The Hubble parameter for above model is expressed as
\begin{equation}\label{34}
H(t)=\dfrac{2nt^{-1+2n}\beta}{t^{2n}\beta+a_{0}},
\end{equation}
where $a_{0},\beta$ are positive constants, $n$ is positive natural number. The bouncing point occur at $t=0$. Scale factor decreases for $t<0$ and increases with $t>0$, expressing contraction and expansion phases respectively.
If $a_{0}$ is substituted zero, a bounce with singularity appears because the Hubble parameter will not be defined at $t=0$. Therefore, we find divergent behaviour of Hubble parameter $H$ and GB invariant $\mathcal{G}$. However, the scale factor keeps on increasing and does not become singular.
The deceleration parameter is defined as follows,
\begin{equation}\label{35}
q(t)=-\dfrac{(-1+2n)t^{-2n}(t^{2n}\beta + a_{0})}{2n\beta}.
\end{equation}
Figure 7 (a) and (b) show variation of scale factor and Hubble parameter for different values of bouncing parameter respectively. Figure 7 (c) shows that universe evolve in accelerated phase of expansion.
\begin{figure}[th!]
\centering \epsfig{file=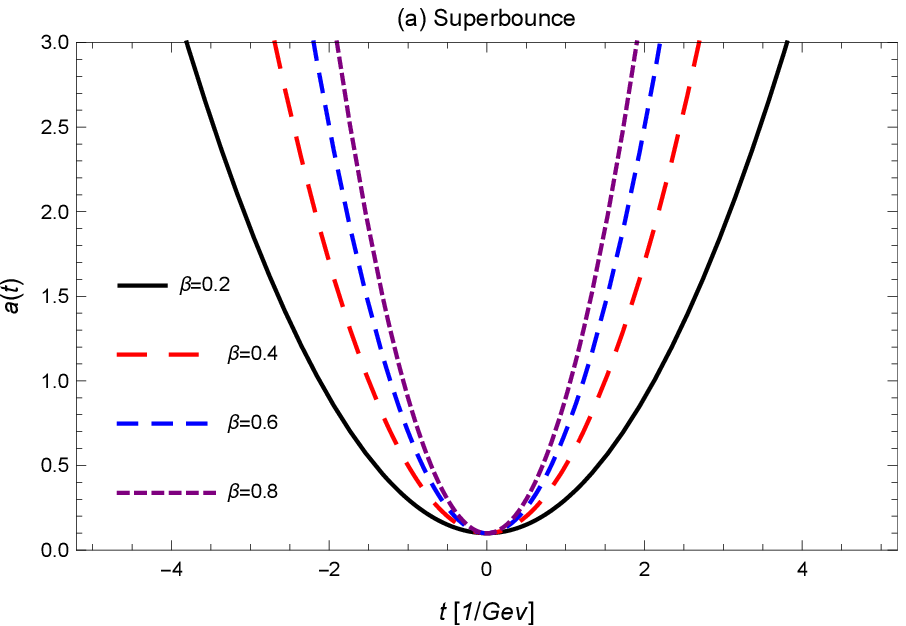, width=.35\linewidth,
height=1.51in}\epsfig{file=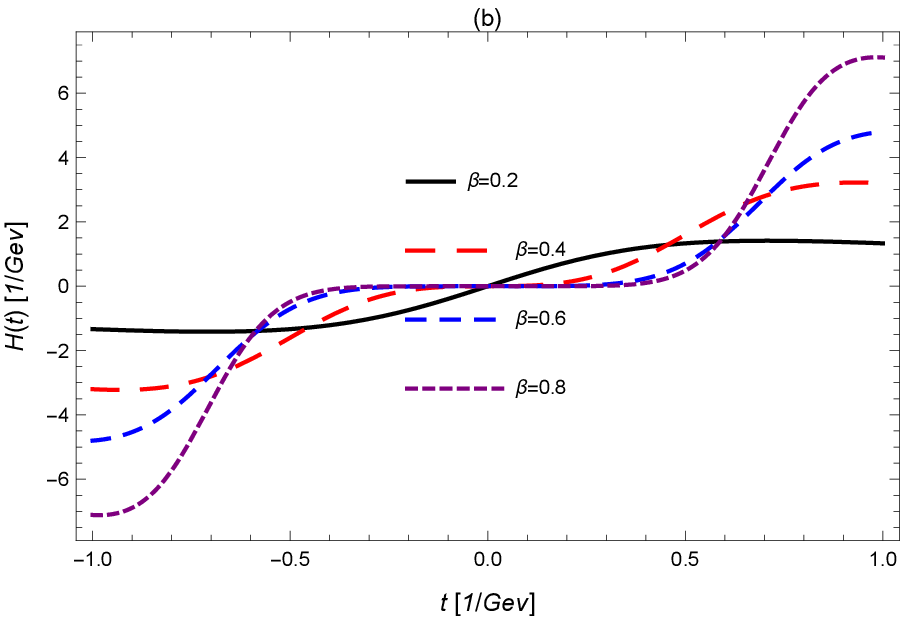, width=.35\linewidth,
height=1.51in}\epsfig{file=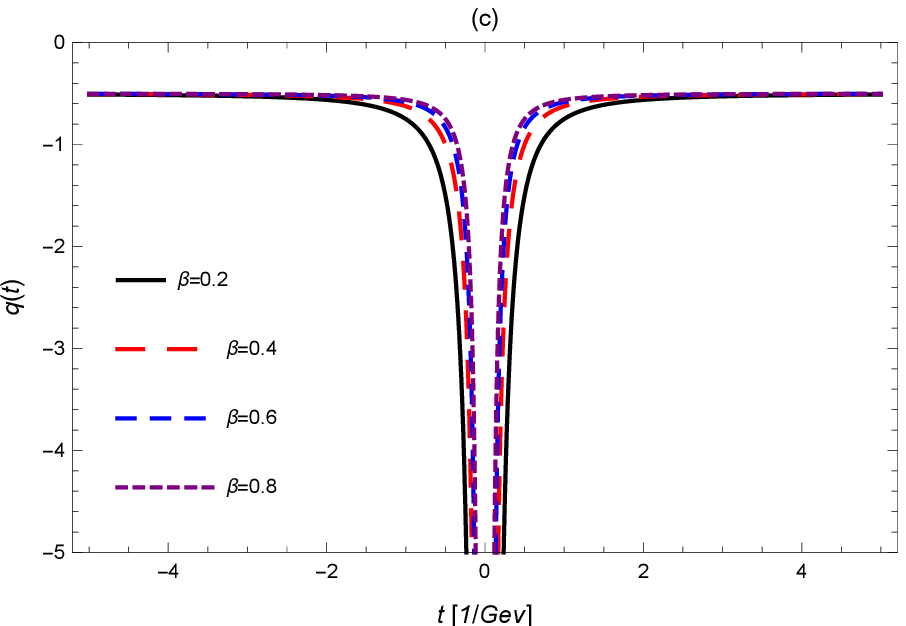, width=.35\linewidth,
height=1.51in}\caption{\label{Fig.7} Evolution of $a(t)$, $H(t)$ and $q(t)$ against $t$ for different values of $\beta$.}
\end{figure}
Now, for superbounce energy density $\rho(t)$, pressure $p(t)$ and EoS parameter $w(t)$ are as follows,
\begin{eqnarray}
\rho(t)&=&\dfrac{96n^{4}t^{-4+8n}\gamma\beta^{4}}{\bigg(t^{2n}\beta+a_{0}\bigg)^{4}}+\dfrac{12M_{p}^{2}n^{2}t^{-2+4n}\beta^{2}}{\bigg(t^{2n}\beta+a_{0}\bigg)^{2}},\\\label{36}
p(t)&=&-\dfrac{96n^{4}t^{-4+8n}\gamma\beta^{4}}{\bigg(t^{2n}\beta+a_{0}\bigg)^{4}}-\dfrac{12M_{p}^{2}n^{2}t^{-2+4n}\beta^{2}}{\bigg(t^{2n}\beta+a_{0}\bigg)^{2}}-2\bigg[M_{p}^{2}+\dfrac{16n^{2}t^{-2+4n}\gamma\beta^{2}}{\bigg(t^{2n}\beta+a_{0}\bigg)^{2}}\bigg]\bigg[-\dfrac{4n^{2}t^{-2+4n}\beta^{2}}{\bigg(t^{2n}\beta+a_{0}\bigg)^{2}}\\\nonumber&+&\dfrac{2n(-1+2n)t^{-2+2n}\beta}{\bigg(t^{2n}\beta+a_{0}\bigg)}\bigg],\\\label{37}
w(t)&=&-1-\dfrac{\bigg[t^{-2n}\bigg(t^{2n}\beta+(1-2n)a_{0}\bigg)\bigg(t^{4n}(M_{p}^{2}t^{2}+16n^{2}\gamma)\beta^{2}+2M_{p}^{2}t^{2+2n}\beta a_{0}+M_{p}^{2}t^{2}a_{0}^{2}\bigg)\bigg]}{3n\beta\bigg(t^{4n}(M_{p}^{2}t^{2}+8n^{2}\gamma)\beta^{2}+2M_{p}^{2}t^{2+2n} \beta a_{0}+M_{p}^{2}t^{2}a_{0}^{2}\bigg)}.\label{38}
\end{eqnarray}

\begin{figure} [th!]
	\epsfig{file=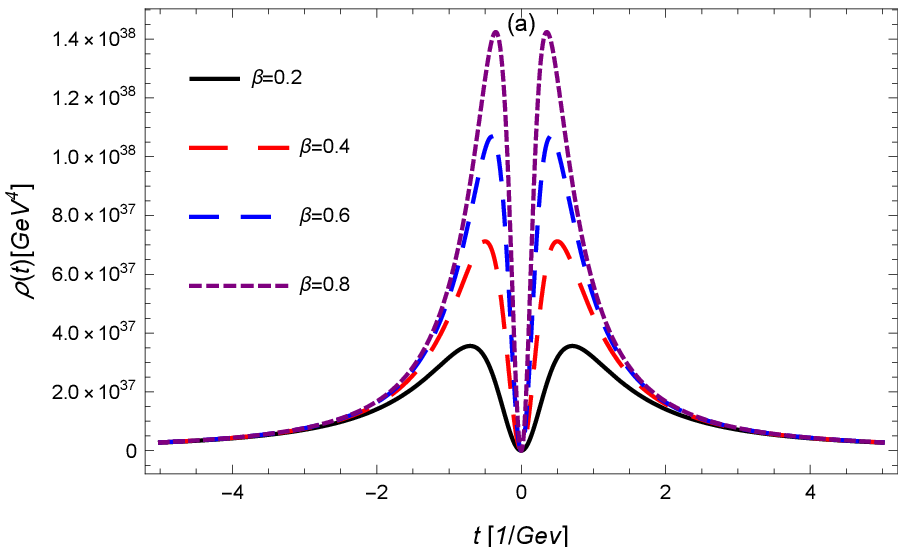,width=0.47\linewidth}\epsfig{file=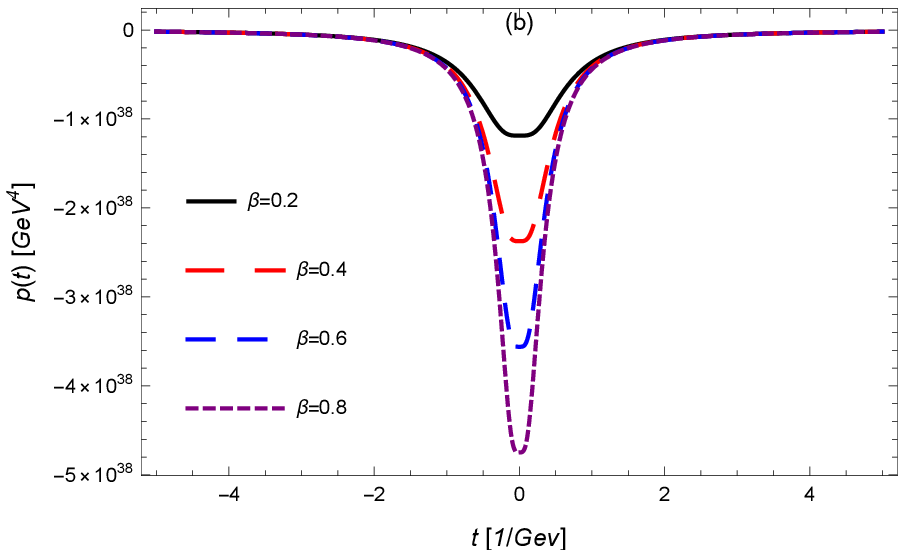,width=0.47\linewidth}
	\caption{The left plot (a) corresponds to the behaviour of $\rho(t)$, whereas right plot (b) depicts the behaviour of $p(t)$. Both graphs are plotted versus $t$ for different values of $\beta$ and $\gamma=2$.}\label{fig8}
\end{figure}

\begin{figure} [th!]
	\epsfig{file=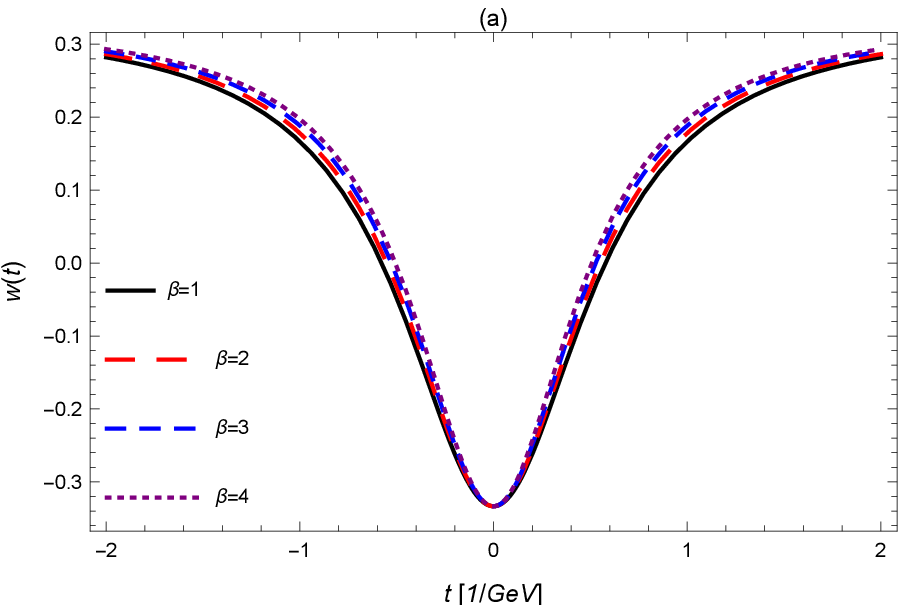,width=0.47\linewidth}\epsfig{file=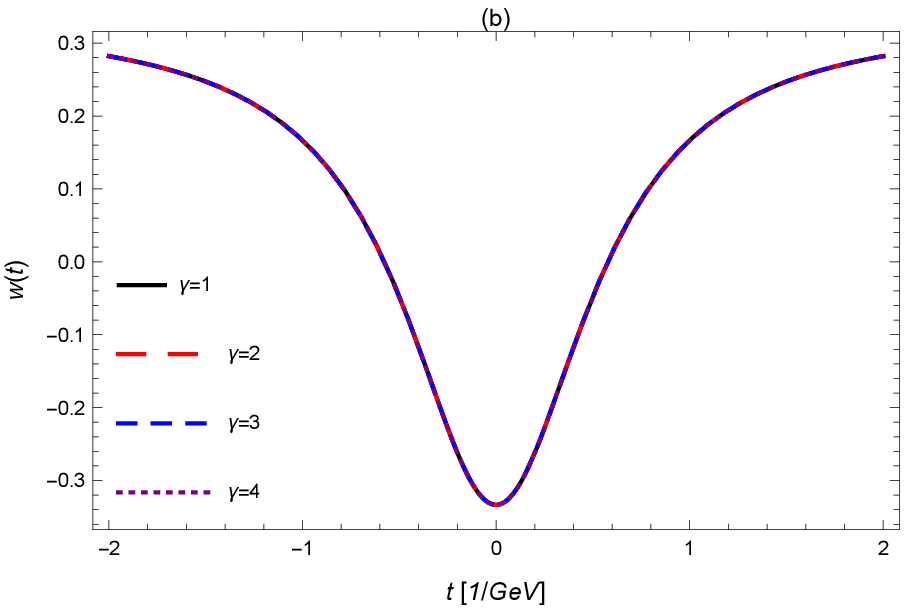,width=0.47\linewidth}
	\caption{The left plot (a) corresponds to the behaviour of $w(t)$  against $t$ when $\gamma=1$ and $\beta$ is varying, whereas plot (b) depicts the behaviour of $w(t)$ against $t$ when $\beta=1$ while $\gamma$ is varying.}\label{fig9}
\end{figure}
From Figure 8(a) it can be observed that $\rho(t)$ is positive but opposite behaviour can be observed from Figure 8(b) for pressure profile. \textcolor{red}{ The plots for EoS parameter $w(t)$ is shown in Figure 9. The EoS parameter shows singularity at $t=0$ (bouncing point). For superbounce, the condition obtained is $\dfrac{8\gamma\beta}{a_{0}}<M_{p}^{2}, \quad n=1$ by constrainting EOS parameter. Figure 9(a) illustrates that, the effects of $\beta$ is quite prominent as compared to $\gamma$ on EoS parameter when plotted against cosmic time.}

\subsection{Oscillatory Bounce}
The oscillatory bounce can be expressed in the form of following function \cite{63}.
\begin{equation}\label{39}
a(t)=sin^{2}(\zeta t).
\end{equation}
It represents cyclic universe followed by self-sustaining, infinite cycles.
Hubble parameter for oscillatory bounce is
\begin{equation}\label{40}
H(t)=2\zeta cot(\zeta t).
\end{equation}
In a cyclic universe, a series of contraction and expansion is experienced. When the scale factor becomes zero singularity appears throughout each cycle. In this case Hubble parameter also become singular. The bounce which takes place at $t=n\pi$ (where $n$ is an integer), shows big bang singularity. This singularity can be vanished by using a non-singular scale factor. At $t=(2n+1)\pi$ second bounce occurs, universe reaches its maximal size. This is how the universe stops expanding and starts to contract, regarded as Big Crunch singularity \cite{77}. Figure 10 (a) and (b) show the behaviour of scale factor and Hubble parameter with increasing bouncing parameter $\zeta=0.2,0.4,0.6,0.8$. It can be seen that oscillatory behaviour of scale factor is experienced. The scale factor and Hubble parameter show symmetric curves.
The deceleration parameter is defined as
\begin{equation}\label{41}
q(t)=-1-\dfrac{1}{2}sec^{2}(\zeta t).
\end{equation}
Figure 10 (c) shows that the range of the deceleration parameter is positive. Therefore, oscillating bouncing models present a deceleration universe for bouncing parameter $\zeta=0.2,0.4,0.6,0.8$. Values of deceleration parameter is greater than $-1$ $(q>-1)$. For $q$ belongs to $[-1,0)$, this implies an accelerated era. Therefore some of the range of $q$ falls in the accelerated phase.

\begin{figure}[th!]
\centering \epsfig{file=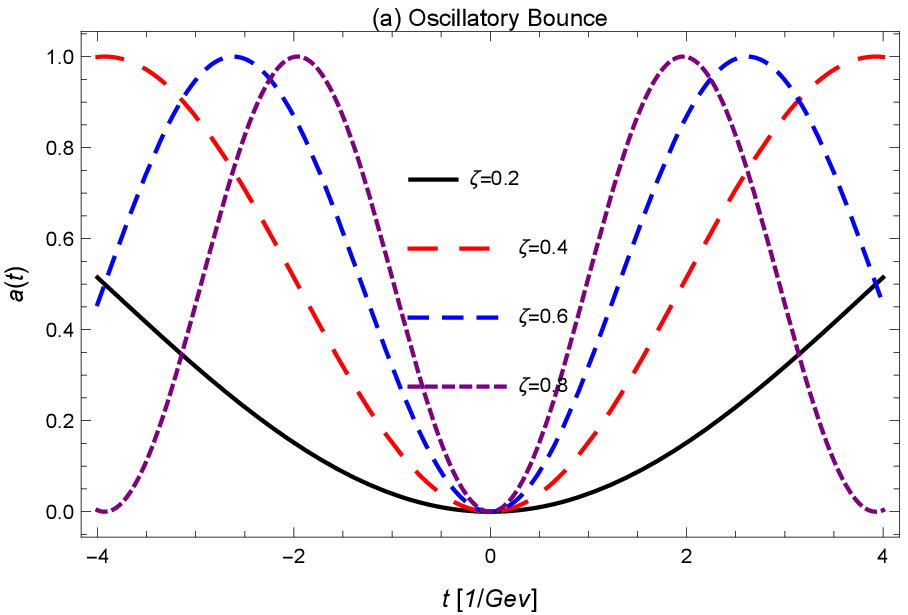, width=.35\linewidth,
height=1.51in}\epsfig{file=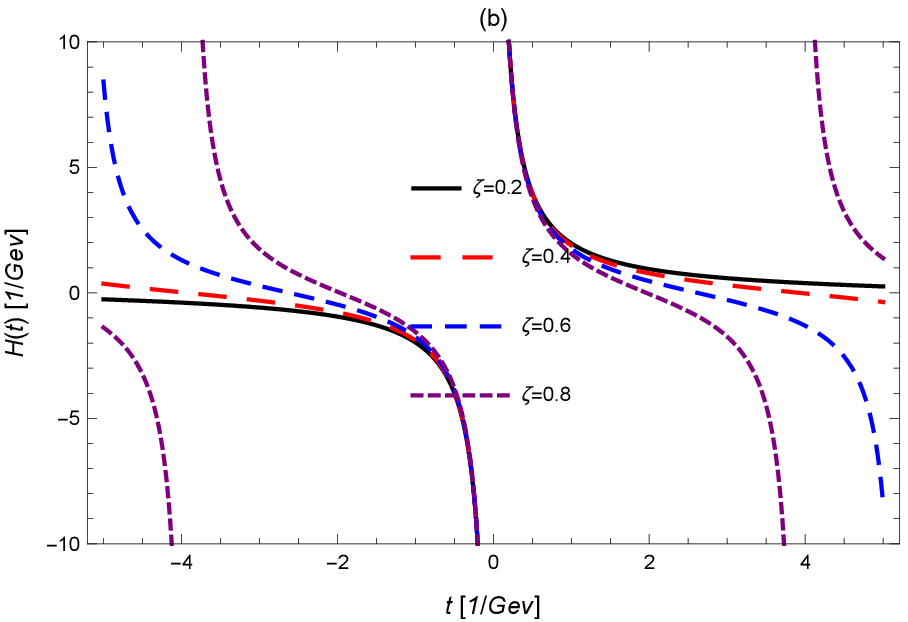, width=.35\linewidth,
height=1.51in}\epsfig{file=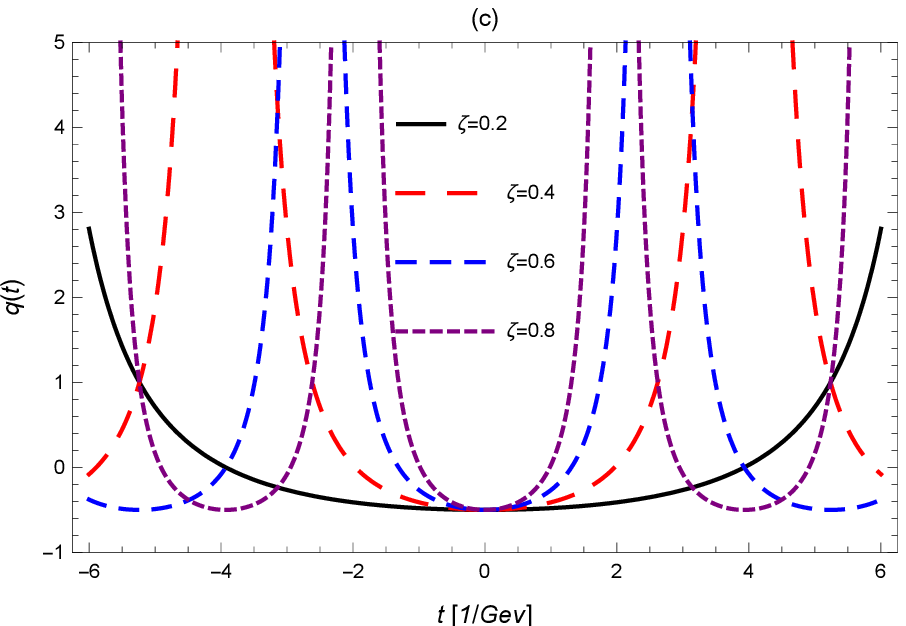, width=.35\linewidth,
height=1.51in}\caption{\label{Fig.10} Evolution of $a(t)$, $H(t)$ and $q(t)$ against $t$ for different values of $\zeta$.}
\end{figure}
Expression for energy density and pressure and EoS parameter are as follows
\begin{eqnarray}\label{42}
\rho(t)&=&12M_{p}^{2}\zeta^{2}cot(\zeta t)^{2}+96\gamma\zeta^{4}cot^{4}(\zeta t),\\\label{43}
p(t)&=&-12 M_{p}^{2}\zeta^{2}cot^{2}(\zeta t)-96\gamma\zeta^{4}cot^{4}(\zeta t)+4\zeta^{4}\bigg(M_{p}^{2}+16\gamma\zeta^{2}cot^{2}(\zeta t)\bigg)cosec^{2}(\zeta t),\\\label{44}
w(t)&=&\frac{\sec^2(\zeta t)\left[M_{p}^2+16\gamma\zeta^2\cot^2(\zeta t)\right]}{3\left[M_{p}^2+8\gamma\zeta^2\cot^2(\zeta t)\right]}-1.\label{44}
\end{eqnarray}
It can be observed from the Eq.(\ref{42}) and Figure 11(a) that energy density is a positive quantity as it largely depends on the bouncing parameter $\zeta$. While Figure 11 (b) shows the behaviour of pressure against cosmic time.
\begin{figure} [th!]
	\epsfig{file=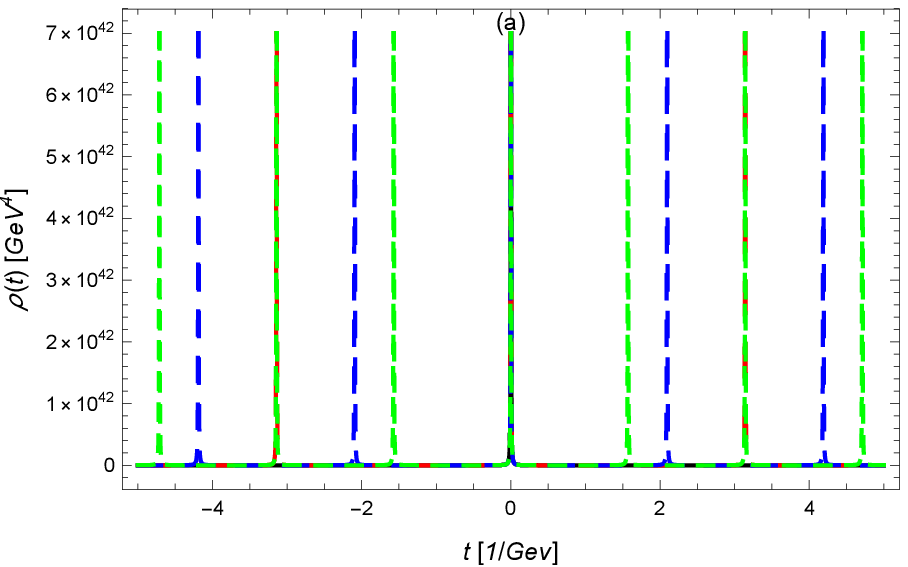,width=0.47\linewidth}\epsfig{file=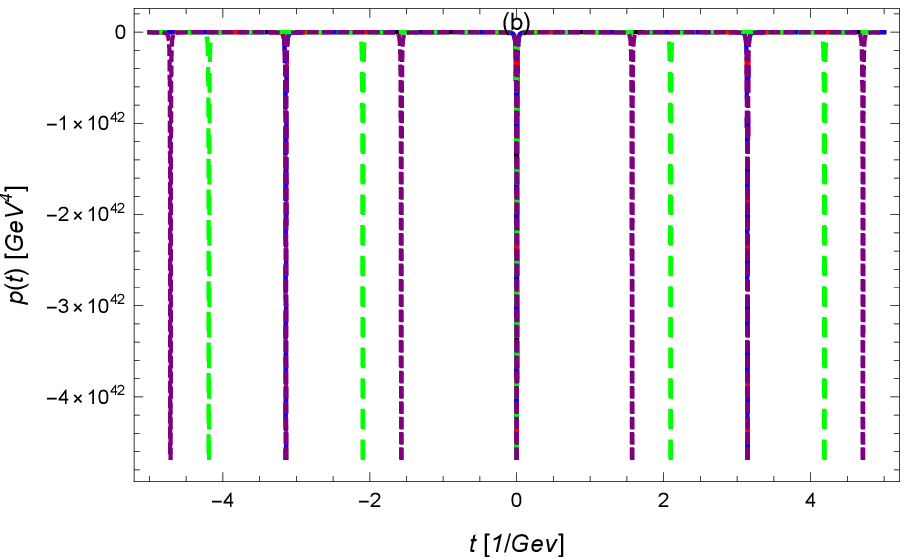,width=0.47\linewidth}
	\caption{The left plot (a) corresponds to the behaviour of $\rho(t)$, whereas right plot (b) depicts the behaviour of $p(t)$. Both graphs are plotted versus $t$ for different values of $\zeta$ which are $\zeta= 0.5, 1, 1.5, 2$.}\label{Fig.11}
\end{figure}
It can be noticed from Figure 12 that the EoS parameter $w(t)$ is greater than -1. This leads us to a non-phantom regime. The variation of EoS parameter $w(t)$, keeping GB coupled parameter $\gamma$ fixed is presented in Figure 12 (a), prominent effects can be experienced. The variation of EoS parameter $w(t)$  when bouncing parameter is constant shown in Figure 12 (b). It can be observed, by varying GB coupled parameter, does not leave an impressive contribution on the EoS parameter $w(t)$.

\begin{figure} [th!]
	\epsfig{file=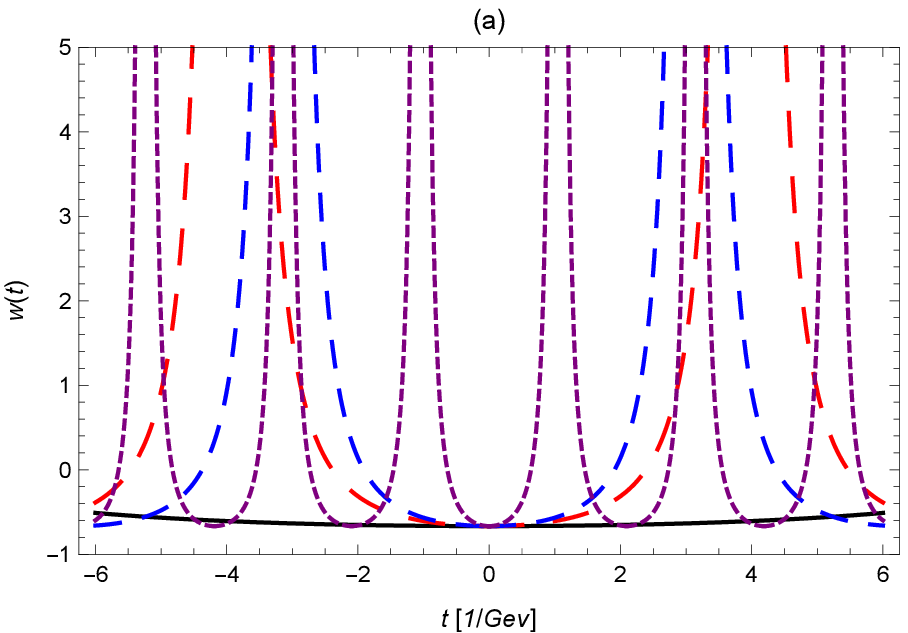,width=0.47\linewidth}\epsfig{file=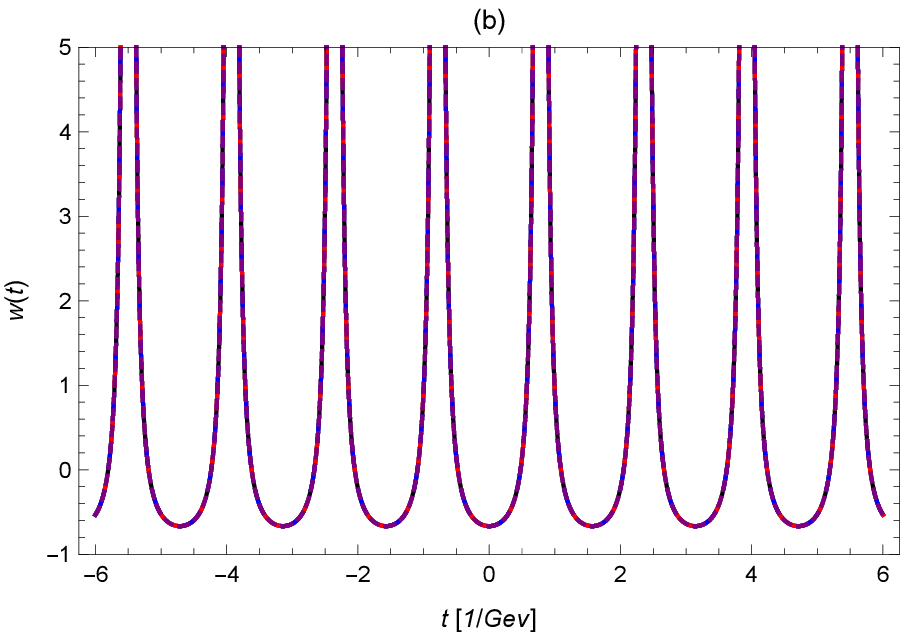,width=0.47\linewidth}
	\caption{The left plot (a) corresponds to the behaviour of $w(t)$ against $t$ when $\gamma = 0$ and $\zeta = 2, 4, 6, 8$ , whereas plot
	 (b) depicts the behaviour of $w(t)$  against $t$ when  $\zeta =1$ and  $\gamma= 2, 4, 6, 8$.}\label{fig12}
\end{figure}

\section{Energy Conditions}
In GR and modified theories of gravitation, when it is impractical to express matter content explicitly, one alternative is to explore energy conditions. Any reasonable matter content will satisfy these conditions. These constraints are not a physical property of the system. Instead, these are mathematically imposed conditions. Moreover, energy conditions gives the criteria about the state of matter and its common properties. It also offers information about all well established non-gravitational fields in physics, being sufficiently fit to eliminate various unphysical solutions of field equations \cite{78}.
For perfect fluid
\begin{equation}\label{45}
T_{\alpha\beta}= (\rho+p)u_{\alpha}u_{\beta}+g_{\alpha\beta}p.
\end{equation}
where $\rho$ is energy density, $p$ is pressure and $u_{\alpha},u_{\beta}$ is four velocity. These constraints are as follows\\

$\bullet$ Weak energy condition (WEC): $\implies$  $\rho(t)\geq0$ and $\rho(t)+p(t)\geq0$.\\

$\bullet$ Null energy condition (NEC): $\implies$  $\rho(t)+p(t)\geq0$.\\

$\bullet$ Dominant energy conditions (DEC): $\implies$ $\rho(t)\geq0$ and $\rho(t)\pm p(t)\geq0$.\\

$\bullet$ Strong energy conditions (SEC): $\implies$ $\rho(t)+3p(t)\geq0$ and $\rho(t)+p(t)\geq0$.\\

\begin{figure}

\epsfig{file=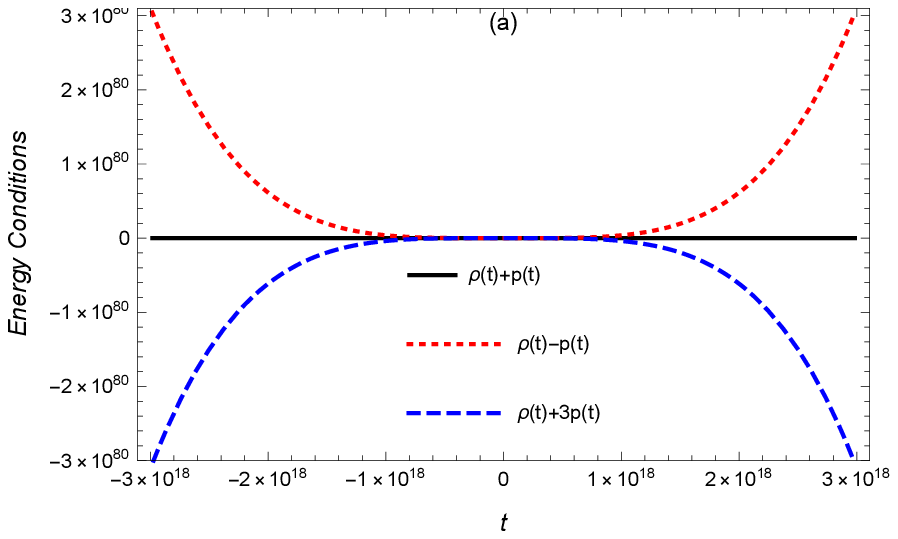,width=0.47\linewidth}\epsfig{file=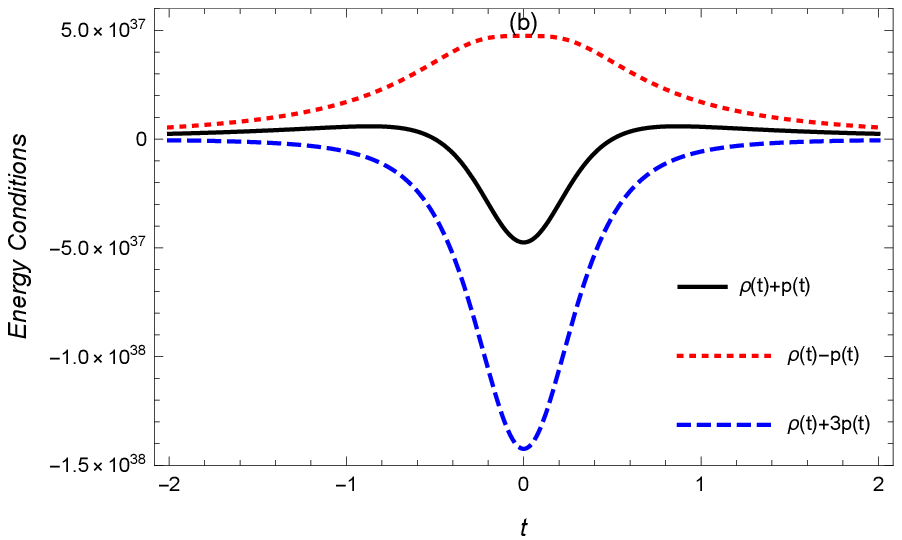,width=0.47\linewidth}
	\caption{The plot (a) corresponds to the energy conditions of Model A, while plot (b) represents the energy conditions of Model B.}\label{fig13}
\end{figure}

\begin{figure}
\epsfig{file=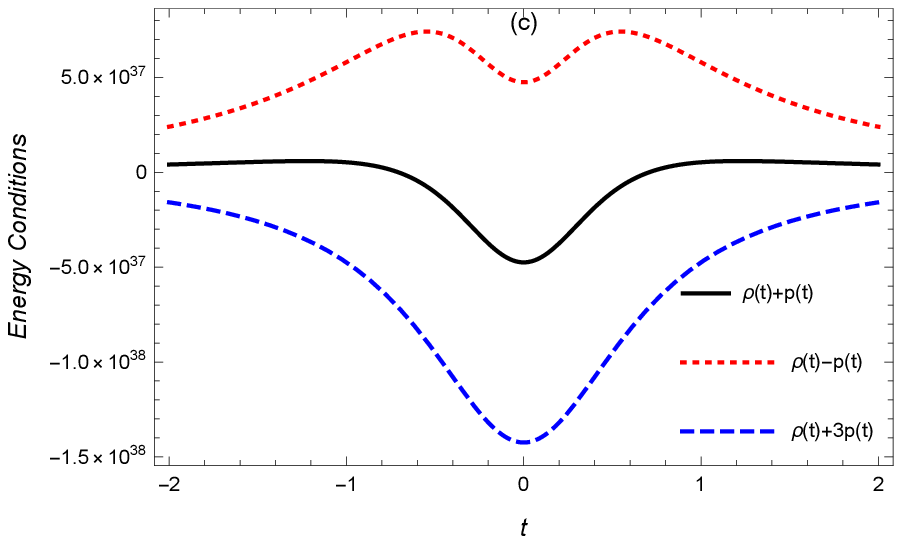,width=0.47\linewidth}\epsfig{file=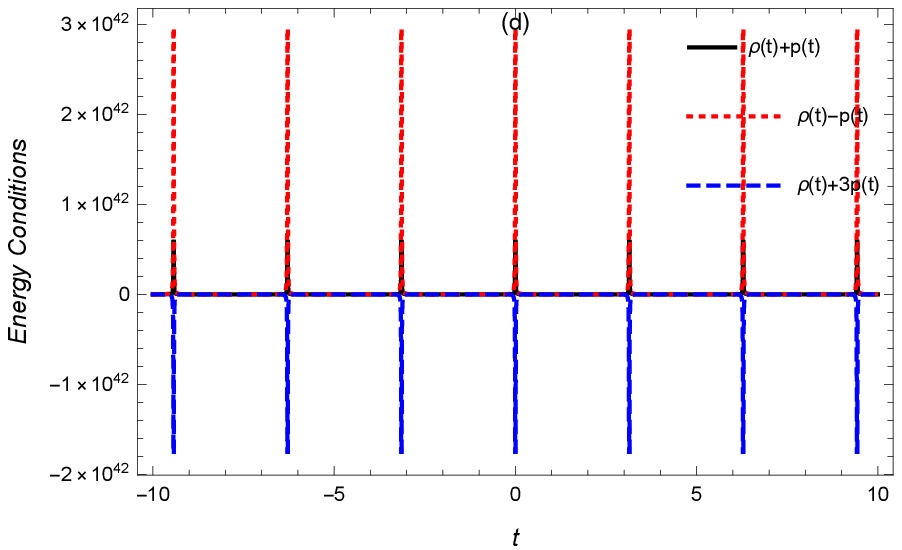,width=0.47\linewidth}
	\caption{The plot (c) corresponds to the energy conditions of Model C, while plot (d) represents the energy conditionsof Model D.}\label{fig14}
\end{figure}

Model A evolves in the phantom region definitely some of the energy conditions are violated. Energy conditions for the symmetric model have been plotted in Figure 13 (a). $\rho$ is positive for all values of cosmic time $t$. There is no singularity near the bouncing epoch ( $t=0$ is bouncing point) for symmetric bounce and it is symmetric in nature. In the symmetric bouncing scenario, near the bouncing point, $\rho+p$ and $\rho+3p$ show negative values, which shows that the model expands in the phantom region. Energy conditions for matter bounce have been plotted in Figure 13 (b). In the present matter bouncing scenario, near the bouncing point, $\rho+p$ and $\rho+3p$ show negative ranges, which take us to the fact that the model exists in the phantom region. Figure 14 shows the energy conditions of the super bounce scenario. It can be observed from Figure 14 (c), weak and strong energy conditions are not satisfied. Violation of energy conditions brings us to the point that, super bounce model also evolves in the phantom era. Energy conditions of the oscillating bouncing model are plotted in Figure 14 (d). It can be seen from Figure 14 (d) Null and dominant energy conditions are satisfied, but strong energy condition is violated. Therefore we can say that this fact leads us to the non-phantom phase.

\section{Cosmography of bouncing models}
The need of the hour is to introduce a model-independent approach to characterize the dark energy behaviour because of dissipation among the cosmological models. This approach only relies on observational assumptions of cosmological facts. The foundation of the standard cosmographic technique is based on the Taylor series expansion of observables. These observables can be compared to data and the results of this technique are free from the EoS parameter. Therefore, cosmography is a very fantastic tool to break the degeneracy between cosmological models and considerably adopted techniques to recognize the dynamics of the universe. Alam et al. \cite{79} founded new cosmological tool $(r,s)$ known as statefinder. Geometrical interpretation of these pairs permits us to specify the properties of dark energy in a model-independent approach. $r$ and $s$ are dimensionless quantities and assembled only from scale factor and its time derivative. In the background of FRW cosmology, deceleration parameter, jerk parameter and snap parameters have been calculated, and results are compatible with observational data \cite{80}.


The taylor series around the scale factor at present era is presented as
\begin{equation}\label{46}
a(t)=a(t_{0})+\dfrac{1}{1!}\dfrac{\partial a(t_{0})}{\partial t}(t-t_{0})+\dfrac{1}{2!}\dfrac{ \partial^{2} a(t_{0})}{\partial^{2} t}(t-t_{0})^{2}+\dfrac{1}{3!}\dfrac{ \partial^{3} a(t_{0})}{\partial^{3} t}(t-t_{0})^{3}+....,
\end{equation}
where $t_{0}$ denotes present cosmic time. The coefficients in the above expansion are known as cosmographic coefficients. These  coefficients offers better geomatrical understanding of universe. A set of these coefficients involved with derivatives of scale factor $a(t)$ for any cosmic time $t$ are expressed as \cite{84}
\begin{equation}\label{47}
H(t)=\dfrac{\dot{a}}{a}, \quad q(t)=-{a(\dfrac{\ddot{a}}{\dot{a}^{2}})}, \quad j(t)=\dfrac{\dddot{a}}{aH^{3}}, \quad s(t)=\dfrac{\ddddot{a}}{aH^{4}}.
\end{equation}
where $j(t), s(t)$ are jerk parameter and snap parameter respectively. $j$ and $s$ are also know as statefinder pair and denoted as $(j,s)$ or $(r,s)$. This pair is considered to be better tool for geometrical understanding of models.
Different combinations of statefinders represents various dark energy models such as presented in Table I.
\begin{table}[ht]
\centering
\begin{tabular}{||c|c|c||}
\hline
Dark Energy Models & j & s \\
\hline
$\Lambda$CDM & 1 & 0 \\
\hline
SCDM & 1 & 1 \\
\hline
HDE & 1 & $\dfrac{2}{3}$ \\
\hline
CG & $>$1 & $<$0 \\
\hline
Quintessence & $<$1 & $>$0 \\
\hline
Matter dominated & 1 & 0.5 \\
\hline
\end{tabular}
\caption{Statefinder pairs for different dark Energy models}
\end{table}

At a very high shift, there is uncertainty in the observational data of redshift $z$. Therefore, the values of these statefinder pairs are not definite. In table II, we have tested bouncing models against these parameters. Firstly, values of these parameters have been calculated at different bouncing point form analytical expressions of these parameters. On account of observations of high redshift $z$ supernova and other observational data sets, the present day value of deceleration parameter, jerk parameter and snap parameter are $q_{0}=-0.81 \pm 0.14$, $j_{0}=2.16^{+0.81}_{-0.75}$ and $s_{0}=-0.22^{+0.18}_{-0.18}$ respectively \cite{85,86,87}. All values of the deceleration parameter are negative, presenting accelerated expansion phase. The cosmographic constraints in Table II is motivated form ref. \cite{88,89}. On the generalizing the behaviour of these parameters, the jerk parameter certainly evolves from huge positive values to an initial epoch, whereas the snap parameter evolves from $-1$ to an initial phase (to $0$) at past times for symmetric bounce. The jerk parameter for matter bounce evolves from large negative values to $-1$, while snap parameter evolves between $0$ and $1$. Futher, jerk parameter for super bounce evolves form large positive values to $0.5$ while snap parameter evolves around $-1$. Moreover, the oscillatory jerk parameter shows decreasing behaviour while the snap parameter evolves from $0$ to $1.5$. For larger value of cosmic time $t$ Model A (symmetric bounce) shows similar behaviour as that of $\Lambda$CDM $(1,0)$ while for Model B and C its statefinder values are$(0,\dfrac{2}{3})$, and $(\dfrac{3}{8},\dfrac{1}{6})$ respectively. For Model D statefinder is $(0,\dfrac{1}{3})$ as cosmic time vanishes. Therefore, we have analyzed the dynamics of cosmographic parameters for all values of cosmic time.\\

\begin{table}[ht]
\centering
\begin{tabular}{||c|c|c|c|c||}
\hline
\rule[-1ex]{0pt}{2.5ex} CC's & Symmetric Bounce & Matter Bounce & Supper Bounce  & Oscillatory Bounce \\
\hline
\rule[-1ex]{0pt}{2.5ex}  & -1.5 & -1 & -1.5  & -0.47945 \\

\rule[-1ex]{0pt}{2.5ex} $q(t)$ & -1.25 & -0.25 & -1.125 & -0.41062 \\

\rule[-1ex]{0pt}{2.5ex}  & -1.1666 & -0.1111 & -1  & -0.2659 \\

\rule[-1ex]{0pt}{2.5ex}  & -1.25 & -0.0625 & -0.9375  & -0.1452 \\
\hline
\rule[-1ex]{0pt}{2.5ex}  & 2.5 & -3 & 1.5  & -0.04109 \\

\rule[-1ex]{0pt}{2.5ex} $j(t)$ & 1.75 & -0.75 & 0.84375  & -0.1787 \\

\rule[-1ex]{0pt}{2.5ex}  & 1.5 & -0.3333 & 0.6666 & -0.4680 \\

\rule[-1ex]{0pt}{2.5ex}  & 1.375 & -0.1875 & 0.5859  & -1.0601 \\
\hline
\rule[-1ex]{0pt}{2.5ex}  & -0.25 & 0.8888 & -0.8333 & 0.35431 \\

\rule[-1ex]{0pt}{2.5ex} $s(t)$ & -0.14285 & 0.7777 & -0.38333  & 0.43148 \\

\rule[-1ex]{0pt}{2.5ex}  & -0.1 & 0.7272 & -0.2222  & 0.638853\\

\rule[-1ex]{0pt}{2.5ex}  & -0.0769 & 0.7037 & -0.1369  & 1.46135 \\
\hline
\end{tabular}
\caption{Variation of $q(t), j(t)$ and $s(t)$ for different values of bouncing parameter.}
\end{table}
In remaining part of this section we wish to study cosmological parameters interms of redshift $z$ for better vision. Here we have expressed Hubble parameter $H(z)$ of each model interms of redshift $z$ by using the relation $z+1=\dfrac{1}{a}$ in equations (\ref{18}),  (\ref{26}), (\ref{33}), (\ref{39}) \cite{90}.
\begin{eqnarray}
H(z)&=&2\sqrt{\lambda}\sqrt{\log(\dfrac{1}{1+z})},\\\label{48}
H(z)&=&\alpha(z+1)\sqrt{1-a_{0}(z+1)^{2}},\\\label{49}
H(z)&=&2n(z+1)^{1/2n}\beta^{1/2n}(1-a_{0}(z+1))^{2n-1/2n},\\\label{50}
H(z)&=&2\zeta \cot(\sin^{-1}(\dfrac{1}{1+z})).\label{51}
\end{eqnarray}
To understand the dynamics of universe, we can find deceleration parameter $q(z)$, jerk parameter $j(z)$, snap parameter $s(z)$ and lerk parameter $l(z)$ from below mentioned expressions \cite{91}
\begin{eqnarray}
q(z)&=&-1+\dfrac{dH(z)}{dz}\dfrac{(1+z)}{H},\\\label{52}
j(z)&=&-q(z)+2q(z)^{2}+(1+z)\dfrac{dq(z)}{dz},\\\label{53}
s(z)&=&-2j(z)-3q(z)j(z)-(1+z)\dfrac{dj(z)}{dz},\\\label{54}
l(z)&=&-3s(z)-4q(z)s(z)-(1+z)\dfrac{ds(z)}{dz}.\label{55}
\end{eqnarray}
The deceleration parameter $q(z)$ for each model represents the accelerating universe. The $q(z)$ shows singular behaviour at $z=0$ except for matter bounce. Also, in symmetric bounce and oscillatory bounce, singularity appears at $z=0$ for jerk parameter $j(z)$. The positive regime can be observed in the case of the snap parameter for each model. The lerk parameter shows negative values in the case of matter bounce, while for all other models positive range can be observed. These parameters also tell us about the past, present and future value of the universe \cite{91}. To evaluate the nature of dark energy models, statefinder pairs $(j,s)$ has been plotted for each bouncing model in Figure 19 (a), (b) and 20 (c), (d). It can be interpreted that at large value of redshift $z$ $(j,s)$ $\rightarrow$ $(1,1)$ for model A and $(j,s)$ $\rightarrow$ $(3,-15)$ for model B.

\begin{figure}[h]
\begin{center}$
\begin{array}{lll}
\includegraphics[width=55mm]{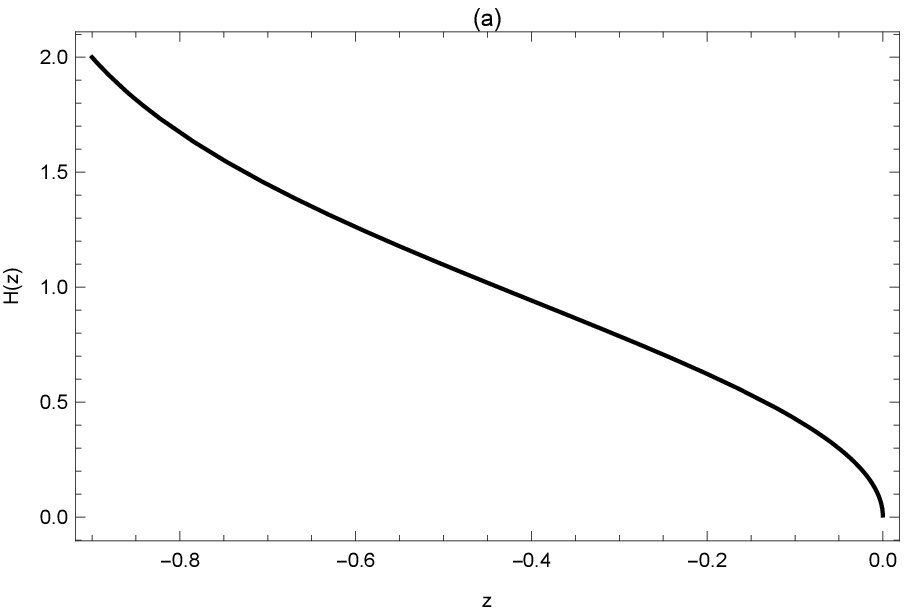}&
\includegraphics[width=55mm]{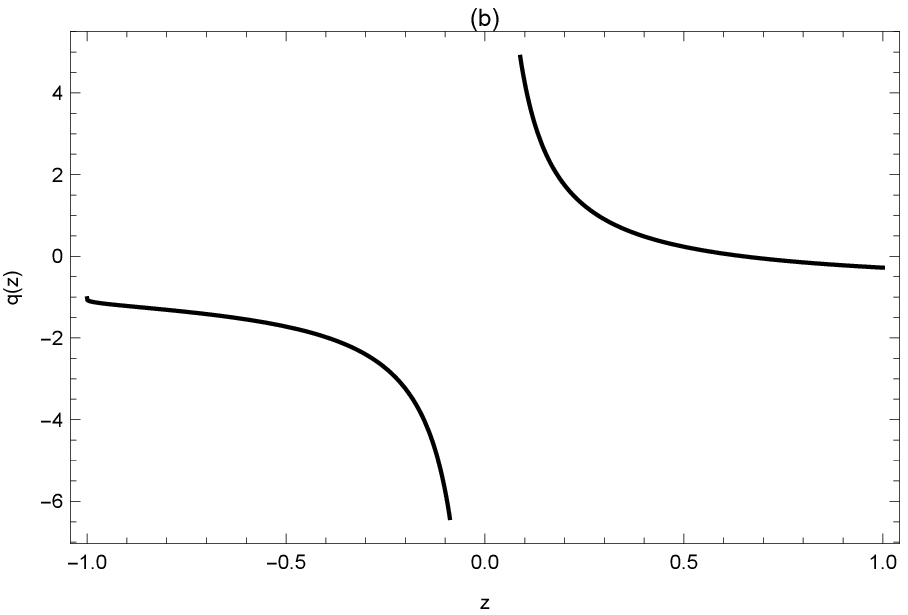}&
\includegraphics[width=55mm]{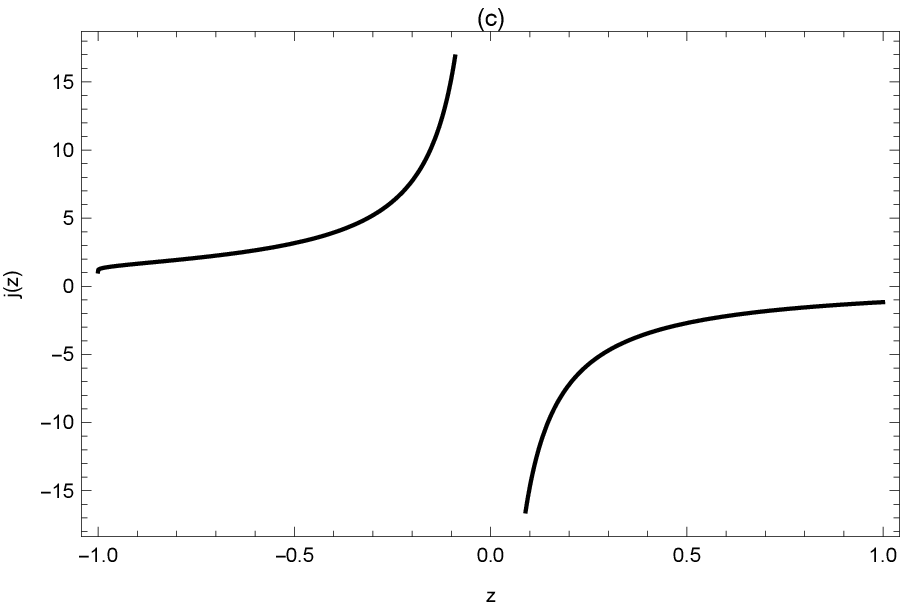}
\end{array}$
\end{center}

\begin{center}$
\begin{array}{rr}
\includegraphics[width=55mm]{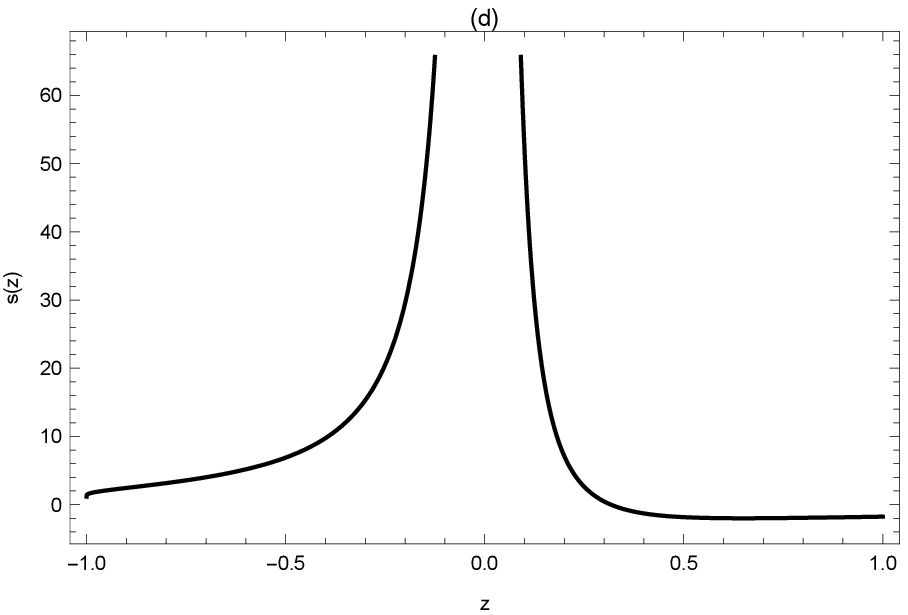}&
\includegraphics[width=55mm]{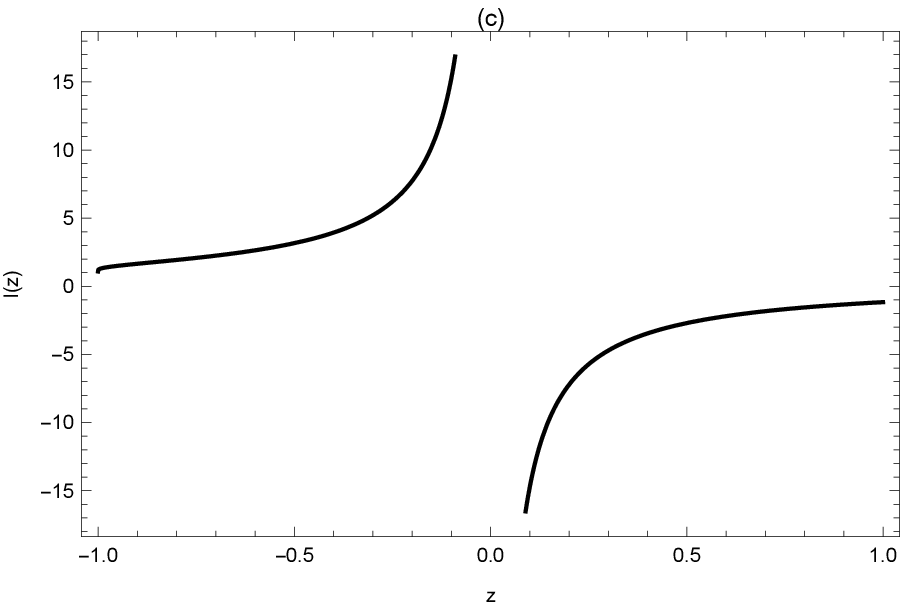}
\end{array}$
\end{center}
\caption{Evolution of $H(z)$, $q(z)$, $j(z)$, $s(z)$ and $l(z)$ for bouncing model A against $z$ when $\lambda=1$.}
\label{pics:blablabla}
\end{figure}

\begin{figure}[h]
\begin{center}$
\begin{array}{lll}
\includegraphics[width=55mm]{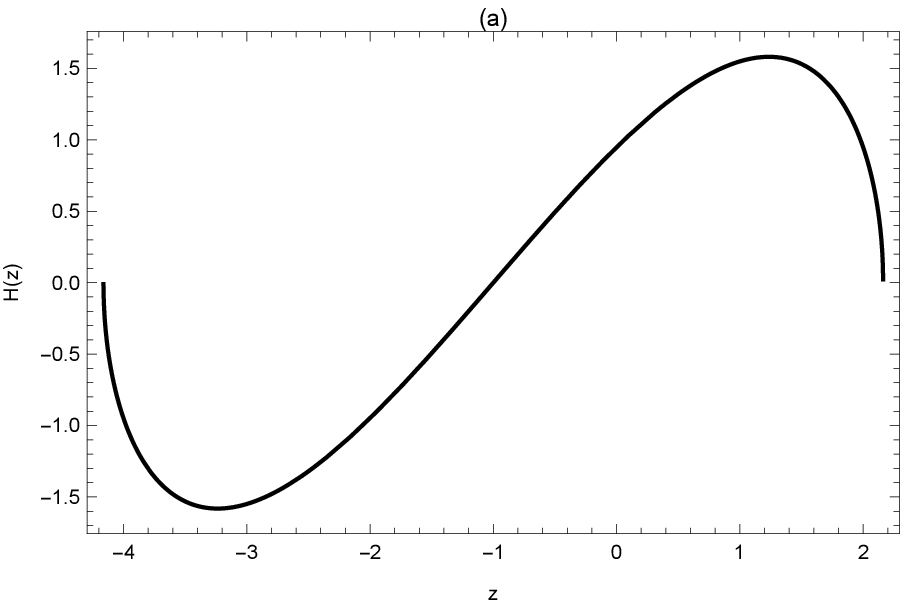}&
\includegraphics[width=55mm]{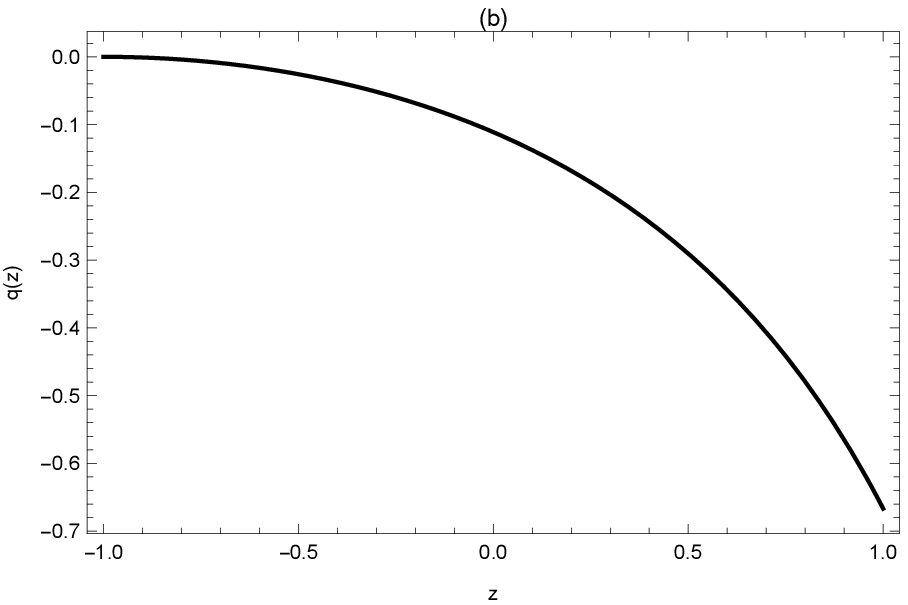}&
\includegraphics[width=55mm]{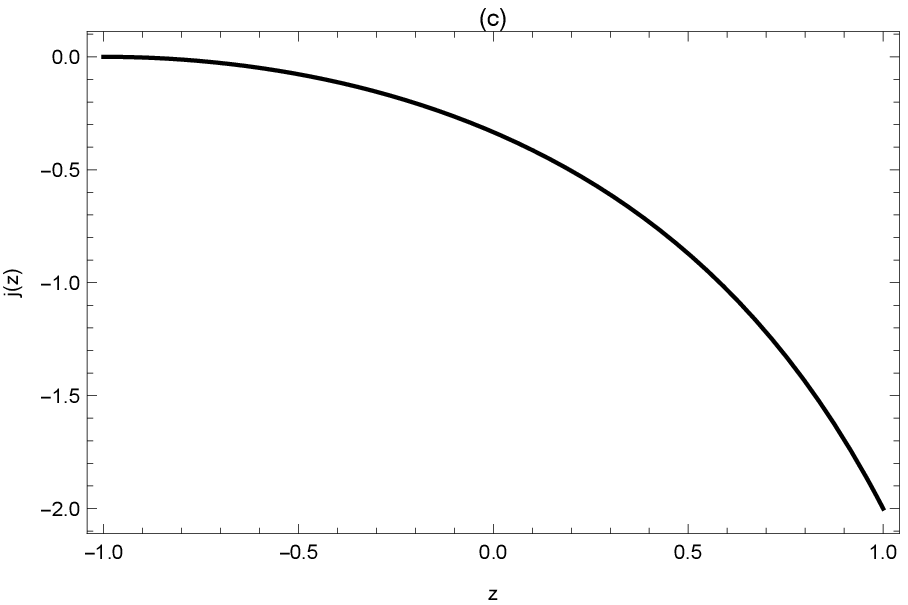}
\end{array}$
\end{center}

\begin{center}$
\begin{array}{rr}
\includegraphics[width=55mm]{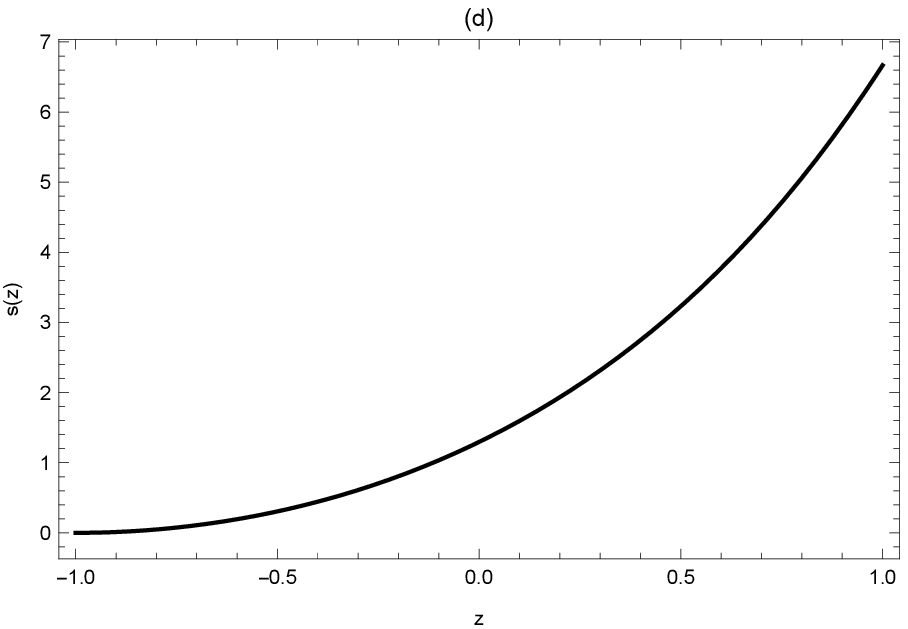}&
\includegraphics[width=55mm]{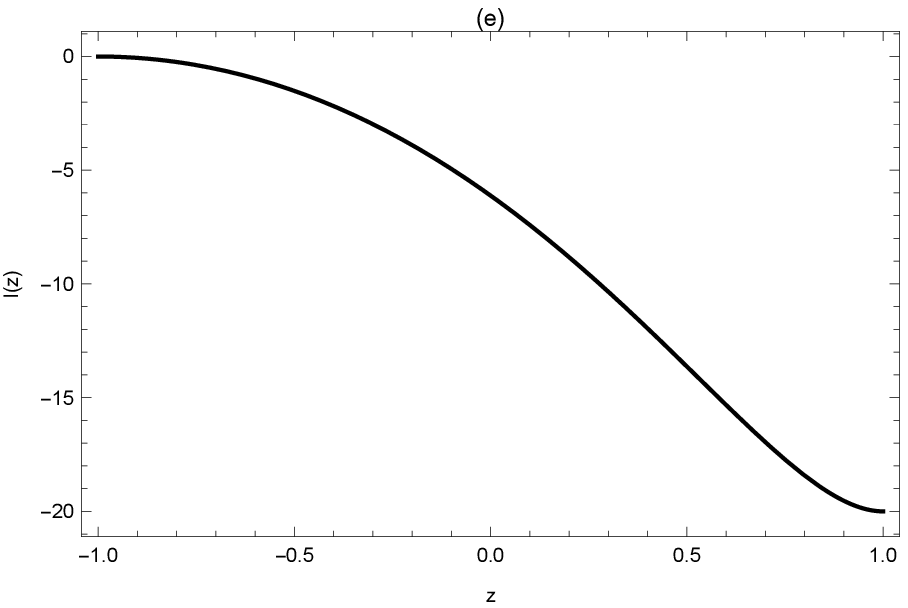}
\end{array}$
\end{center}
\caption{Evolution of $H(z)$, $q(z)$, $j(z)$, $s(z)$ and $l(z)$ for bouncing model B against $z$ when $a_{0}=0.1$ and $\alpha=1$.}
\label{pics:blablabla}
\end{figure}

\begin{figure}[h]
\begin{center}$
\begin{array}{lll}
\includegraphics[width=55mm]{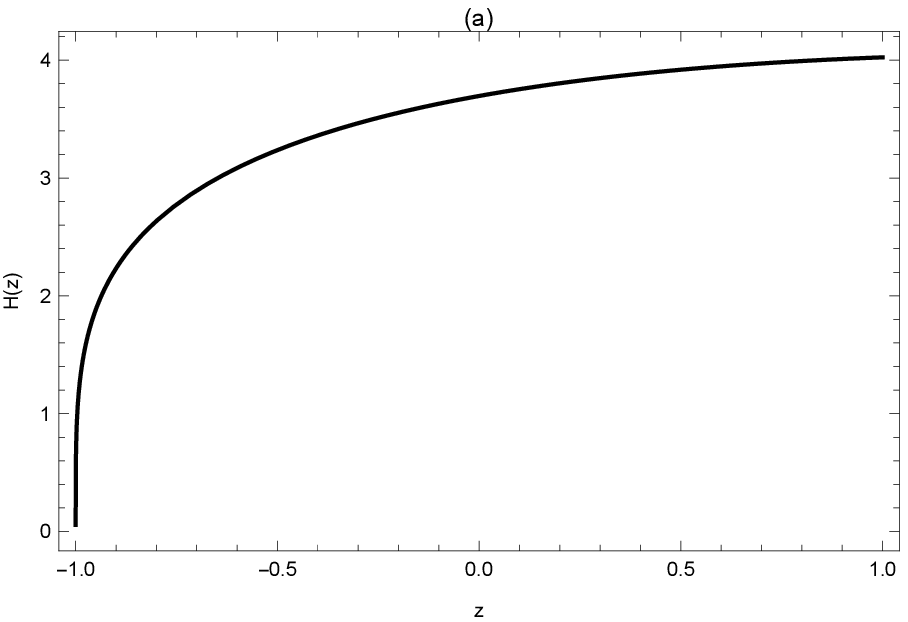}&
\includegraphics[width=55mm]{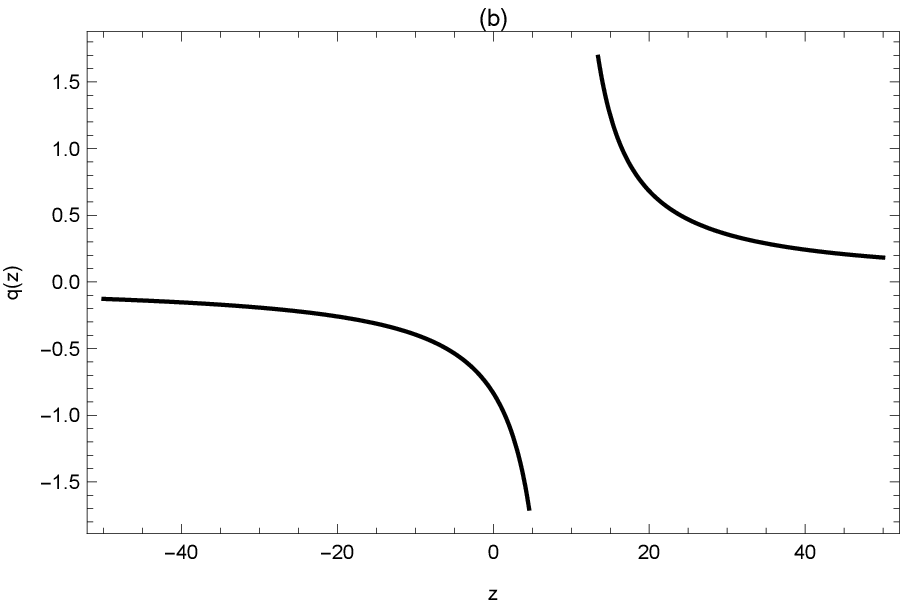}&
\includegraphics[width=55mm]{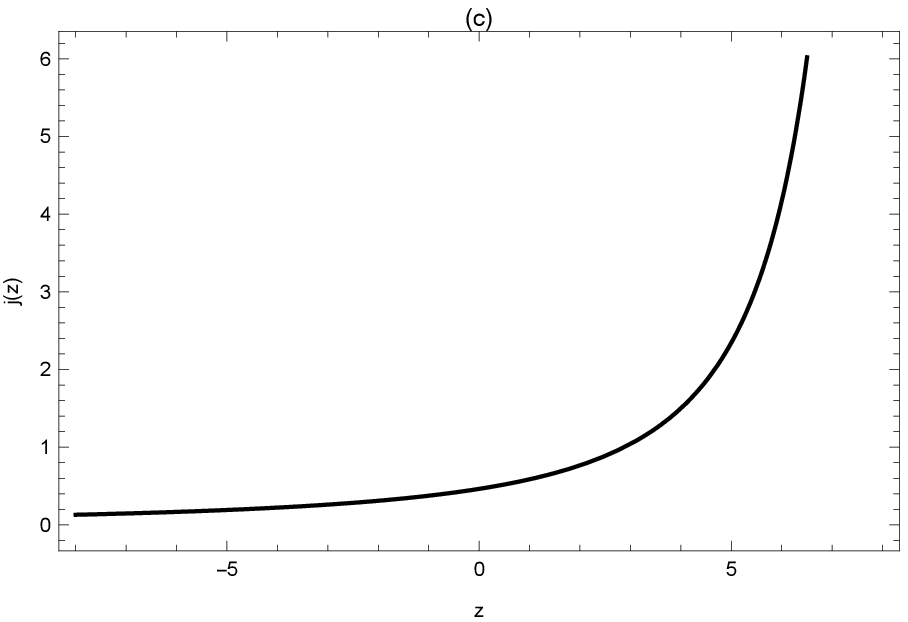}
\end{array}$
\end{center}

\begin{center}$
\begin{array}{rr}
\includegraphics[width=55mm]{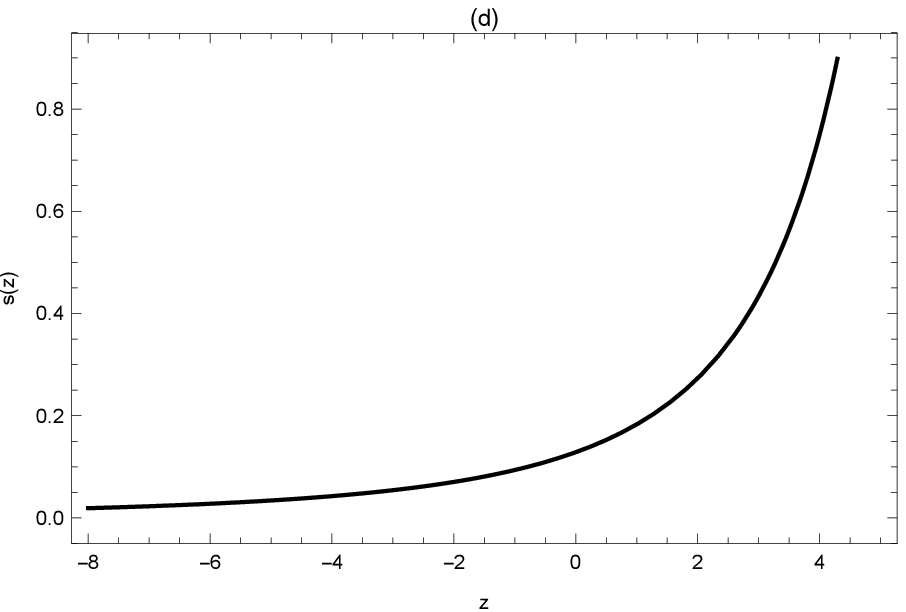}&
\includegraphics[width=55mm]{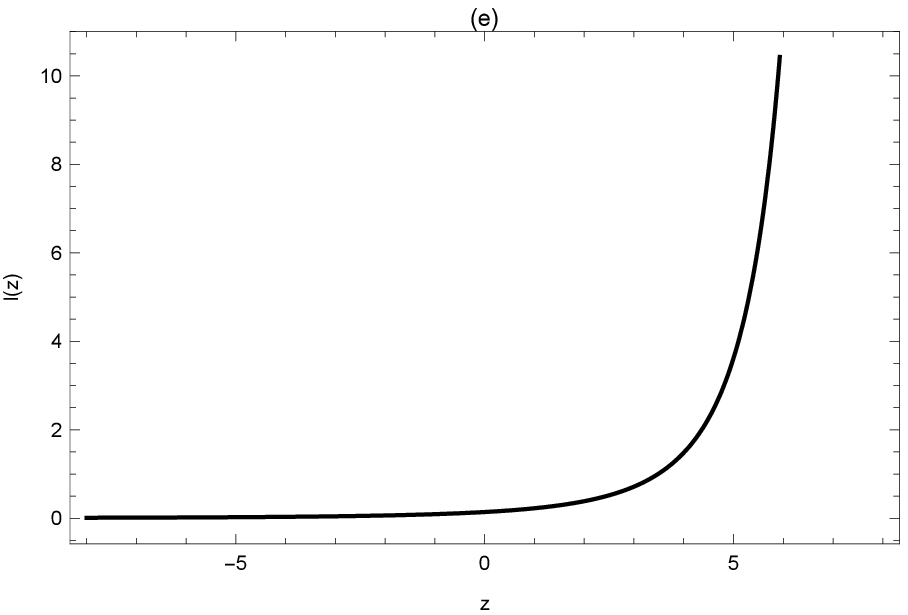}
\end{array}$
\end{center}
\caption{Evolution of $H(z)$, $q(z)$, $j(z)$, $s(z)$ and $l(z)$ for bouncing model C against $z$ when $\beta=1$, $n=2$ and $a_{0}=0.1$.}
\label{pics:blablabla}
\end{figure}

\begin{figure}[h]
\begin{center}$
\begin{array}{lll}
\includegraphics[width=55mm]{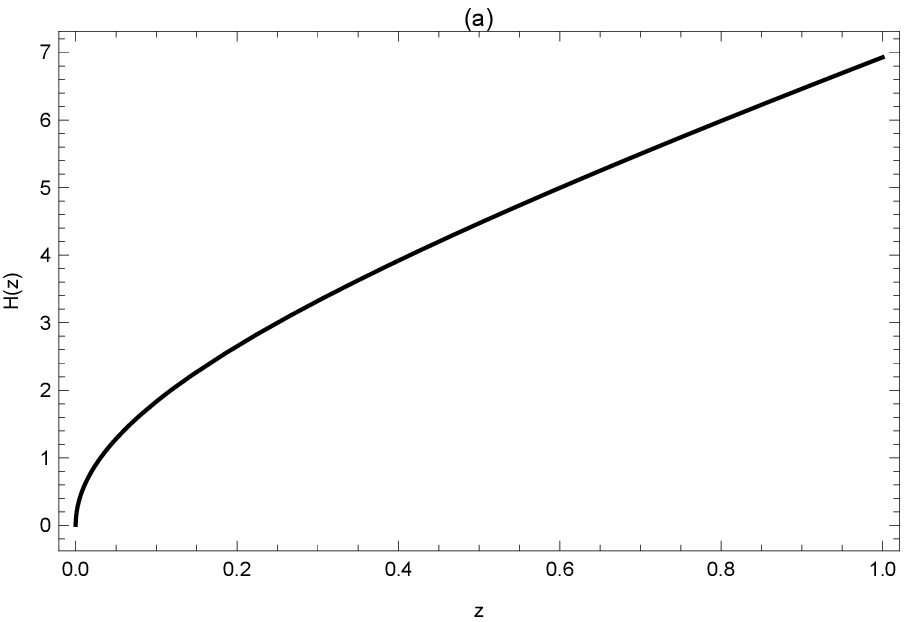}&
\includegraphics[width=55mm]{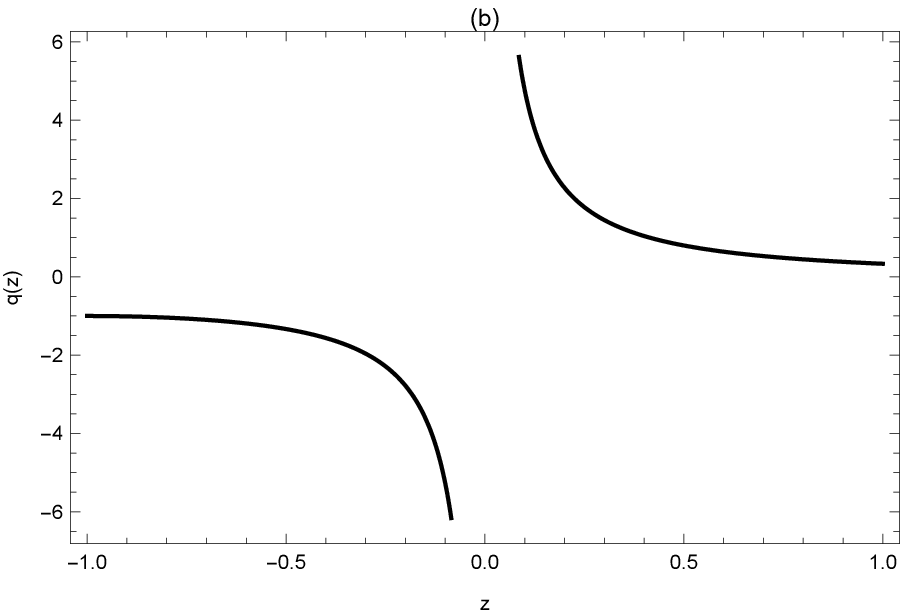}&
\includegraphics[width=55mm]{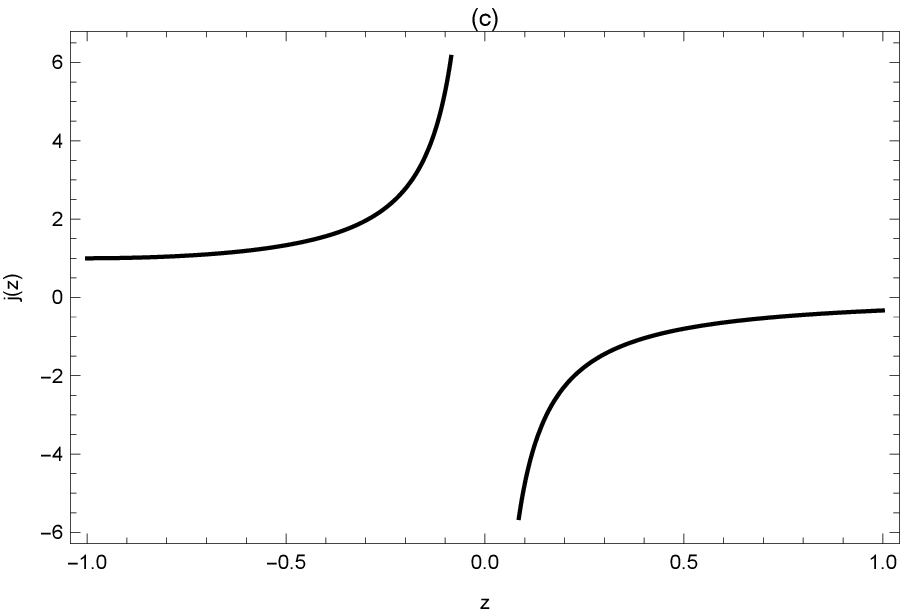}
\end{array}$
\end{center}

\begin{center}$
\begin{array}{rr}
\includegraphics[width=55mm]{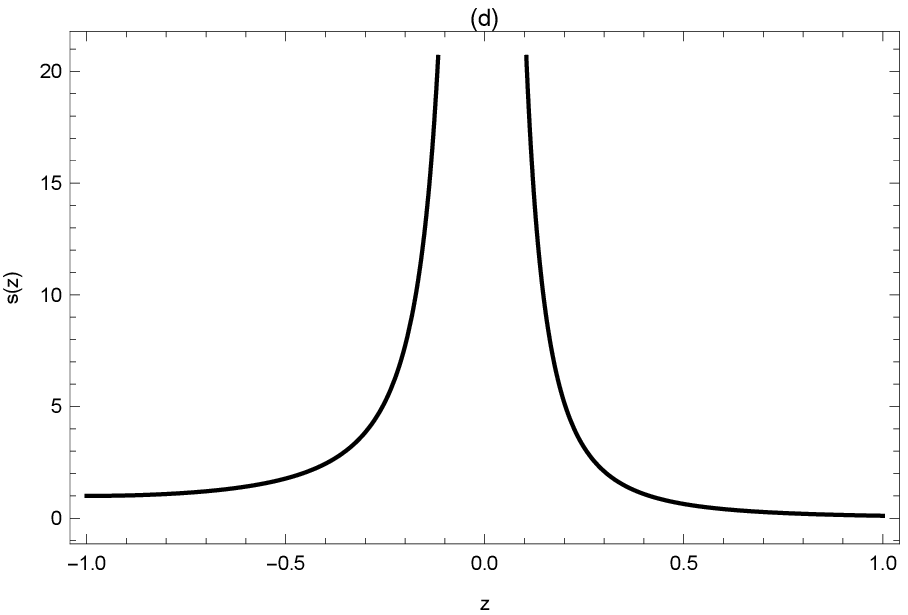}&
\includegraphics[width=55mm]{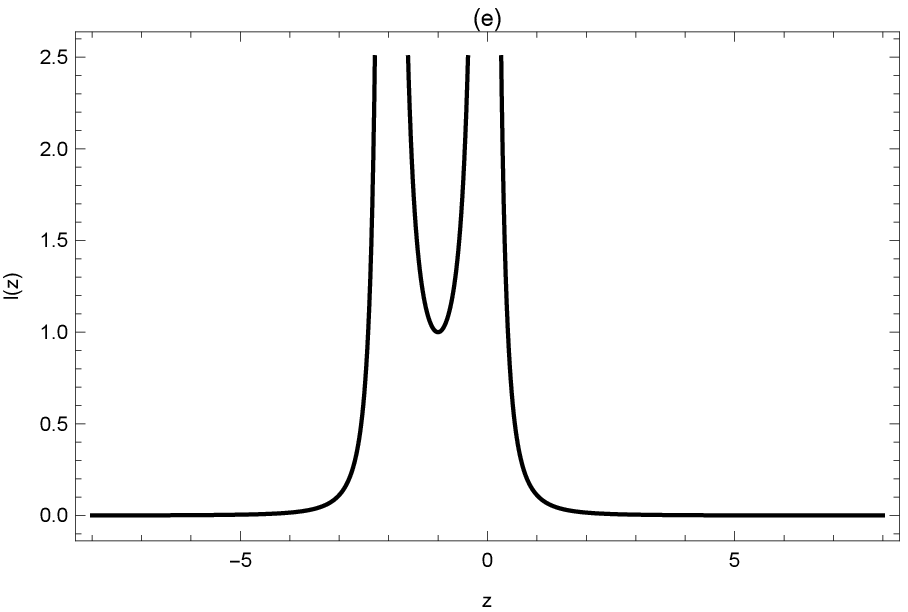}
\end{array}$
\end{center}
\caption{Evolution of $H(z)$, $q(z)$, $j(z)$, $s(z)$ and $l(z)$ for bouncing model D against $z$ when $\zeta=1$.}
\label{pics:blablabla}
\end{figure}

\begin{figure}[th!]
\centering
\epsfig{file=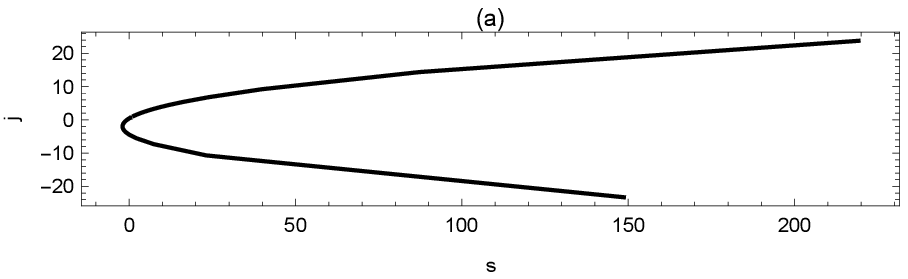, width=.45\linewidth, height=2.2in}
\epsfig{file=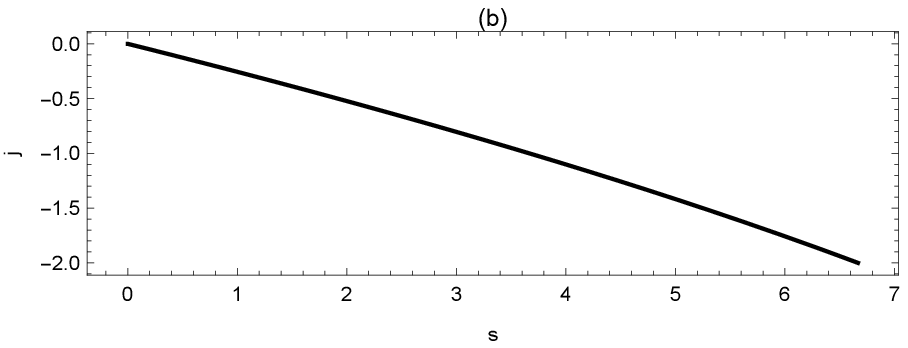, width=.45\linewidth, height=2.2in}
\caption{Left plot shows evolution of $(j,s)$ for model A when $\lambda=1$ while, right plot shows evolution of $(j,s)$ for Model B when $\alpha=1$.} \label{fig10}
\end{figure}
\begin{figure}[th!]
\centering
\epsfig{file=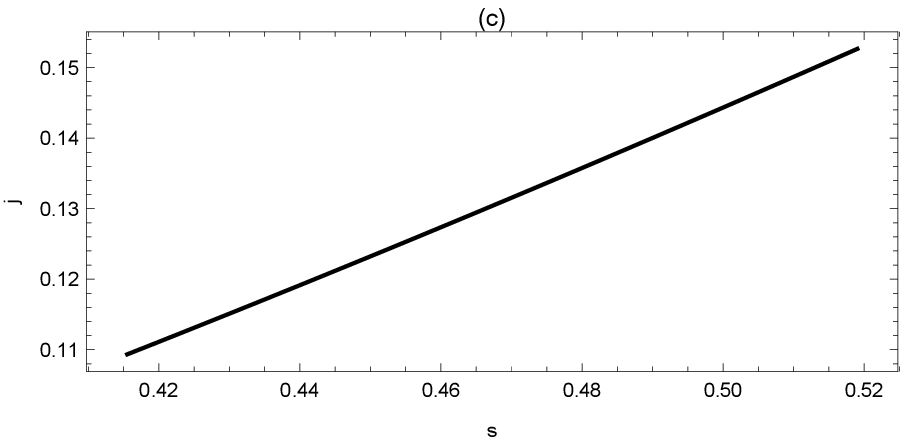, width=.45\linewidth, height=2.2in}
\epsfig{file=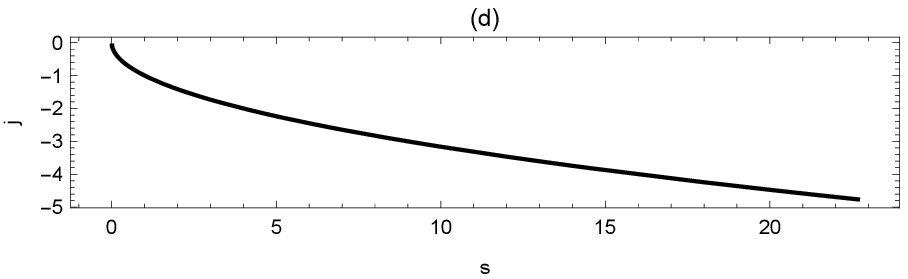, width=.45\linewidth, height=2.2in}
\caption{Left plot shows evolution of $(j,s)$ for model C when $n=2$, $a_{0}=0.1$ and $\beta=1$ while, right plot shows evolution of $(j,s)$ for Model D when $\zeta=1$.} \label{fig10}
\end{figure}

\section{Stability Analysis}
In this section, stability of bouncing models in 4D EGB gravity is discussed by squared speed of sound method. Squared speed of sound is denoted by $C_s^{2}$ and defined as $C_s^{2}=\dfrac{dp}{dz}\times\dfrac{dz}{d\rho}$. In mechanically and thermodynamically stable system, squared speed of sound should give non-negative values. Therefore above bouncing models would be called stable for positive values of Squared speed of sound $C_s^{2}$. For mechanical stability analysis, squared speed of sound $C_s^{2}$ should remain between zero and one.\\
The expressions of squared speed of sound $C_s^{2}$ in terms of redshift for Model A, B, C and D are as follows
\begin{eqnarray}\nonumber
C_s^{2}(z)&=&\dfrac{192\gamma\lambda^{2}\log(\dfrac{1}{1+z})}{1+z}-\dfrac{1+z}{12M_{p}^{2}\lambda}\bigg[\dfrac{12M_{p}^{2}\lambda}{1+z}-\dfrac{32\gamma\lambda^{3/2}}{(1+z)^{2}\sqrt{\log(\dfrac{1}{1+z})}}\\\label{56}
&+&\dfrac{192\gamma\alpha^{2}\log(\dfrac{1}{1+z})}{1+z}+\dfrac{\sqrt{\alpha}(M^{2}_{p}+16\gamma\alpha\log\dfrac{1}{1+z})}{(1+z)^{2}\log(\dfrac{1}{1+z})^{3/2}}-\dfrac{2\sqrt{\alpha}(M_{p}^{2}+16\gamma\alpha\log(\dfrac{1}{1+z}))}{(1+z)^{2}\sqrt{\log(\dfrac{1}{1+z})}}\bigg],\\\nonumber
C_s^{2}(z)&=& 6 M_{p}^{2}(1+z)\alpha^{2} (1-(1+z)^{2} a_{0})-24(1+z)^{5}\gamma\alpha^{4}a_{0}(1-(1+z)^{2}a_{0})\\\nonumber
&+&24(1+z)^{3}\gamma\alpha^{4}(-1+(1 + z)^{2}a_{0})^{2} + \dfrac{1}{3}M_{p}^{2}(1 + z)^{4}\alpha^{3}a_{0}\bigg[\dfrac{1}{(1-(1 + z)^{2} a_{0})^{3/2}}\\\nonumber
&\times&-3M_{p}^{2}(1 + z)^{2}\alpha a_{0}+3M_{p}^2\alpha(1-(1 + z)^{2}a_{0})-12(1+z)^{4} \gamma\alpha^{3}a_{0}(1\\\nonumber
&-&(1 + z)^{2}a_{0})+12(1+z)^{2} \gamma\alpha^{3}(-1+(1 + z)^{2} a_{0})^{2} +\dfrac{(8\gamma(\alpha-2(1 + z)^{2}\alpha a_{0})^{2})}{\sqrt{1-(1-z)^{2}a_{0}}}\\\label{57}
&-&(a_{0}(-3+2(1+z)^{2}a_{0})(-M_{p}^{2}-4(1 + z)^{2}\gamma\alpha^{2}+4(1 + z)^{4}\gamma\alpha^{2} a_{0}))\bigg],\\\nonumber
C_s^{2}(z)&=&-\dfrac{1}{6n}\bigg[72 M_{p}^{2} n^{2} (1 + z)^{-1 + 1/n} \beta^{1/n}(1 - (1 + z) a_{0})^{2-1/n}+16n^{3}(1+z)^{-3 + 5/2n}\\\nonumber
&\times&\gamma\beta^{5/2n}(1-(1+z)a_{0})^{3}- \dfrac{5}{2n} (M_{p}- 2 M_{p} n (1 + z) a_{0})^{2}+6(-1 +2n)\\\nonumber
&\times&(1 + z)^{-2 + 1/2n} \beta^{1/2n}(1-(1+z)a_{0})^{-1-3/2n}(-M_{p}^{2}(1-(1+z)a_{0})^{1/n})\\\label{58}
&-&16 n^{2}(1 + z)^{1/n}\gamma\beta^{1/n}(-1+(1+z)a_{0})^{2}\bigg],\\\nonumber
C_s^{2}(z)&=&\frac{1}{6 \zeta  M_{p}^2 (z+1)^{3} \left(\frac{z (z+2)}{(z+1)^2}\right)^{3/2}}\bigg[-144 \zeta ^3 M_{p}^{4} z (z+2) \sqrt{\frac{z (z+2)}{(z+1)^{2}}}-M_{p}^{2}(z+1) \\\label{59}
&\times& \bigg(768 \gamma \zeta ^4 z^{3}+2304 \gamma  \zeta ^{4} z^{2}+1536 \gamma  \zeta ^{4} z-1\bigg)+16 \gamma \zeta ^{2} z \left(z^{2}+3 z+2\right)\bigg].
\end{eqnarray}
Figures 21 (a) and (b) show stability analysis for Model A (symmetric bounce) and Model B (Matter bounce). It can be seen from Figure 21 (a) for different values of bouncing parameter, squared speed of sound show negative values which predicts unstable behaviour of model A. Stable behaviour of matter bounce model can be observed from Figure 21 (b). When values of bouncing parameter $\alpha$ increases, squared speed of sound $C_s^{2}$ gives positive values. While opposite results can be observed for matter bounce scenario in $f(Q, T)$ gravity \cite{90}. Model C and Model D does not satisfy stability conditions. The GB coupled parameter $\gamma$ does not contribute much to the dynamics of stability analysis. It will only show its contribution when it is taken to be very high.
\begin{figure} [th!]
	\epsfig{file=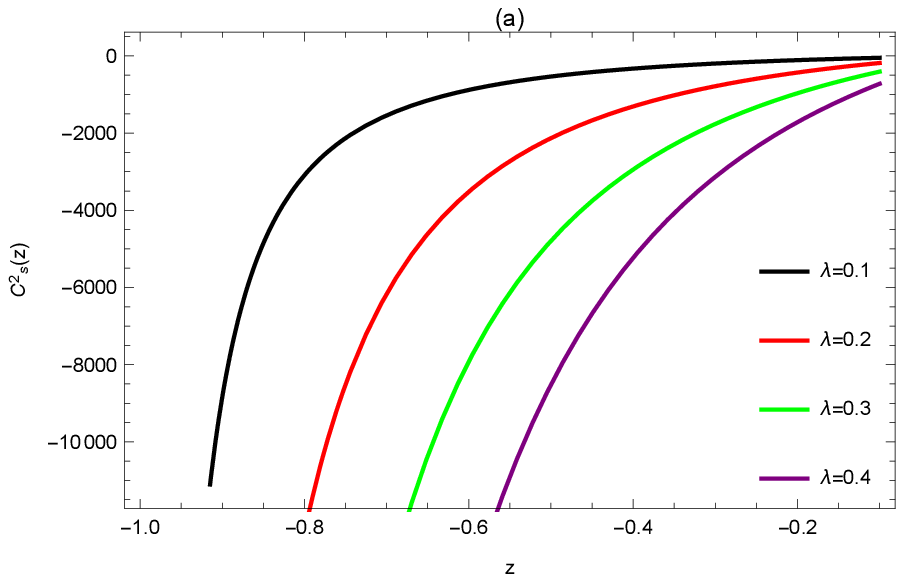,width=0.47\linewidth}\epsfig{file=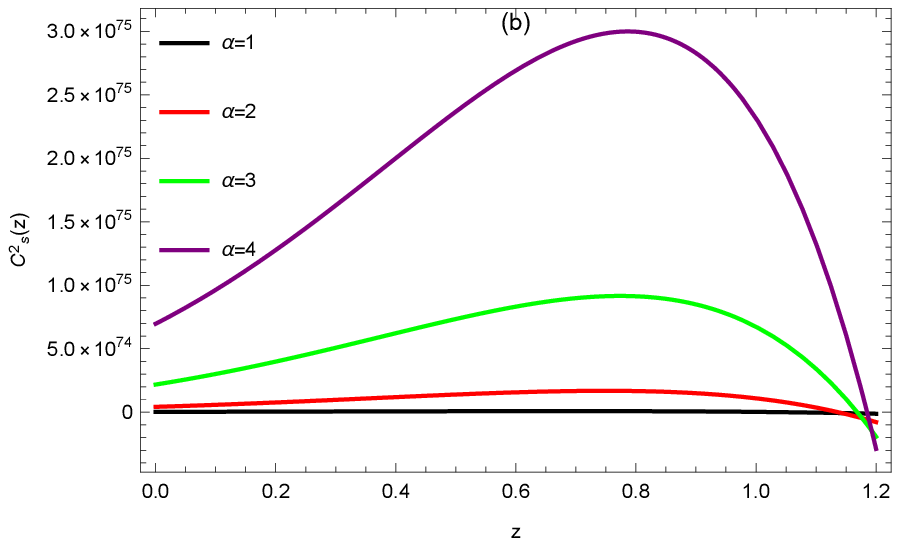,width=0.47\linewidth}
	\caption{ The plot (a) corresponds to the stability analysis of model A for different values of $\lambda$ while $\gamma=2$ whereas plot (b) represents the stability analysis of model B for different values of $\alpha$ while is $\gamma=2$ and $a_{0}=0.1$.}\label{fig14}
\end{figure}
\begin{figure} [th!]
	\epsfig{file=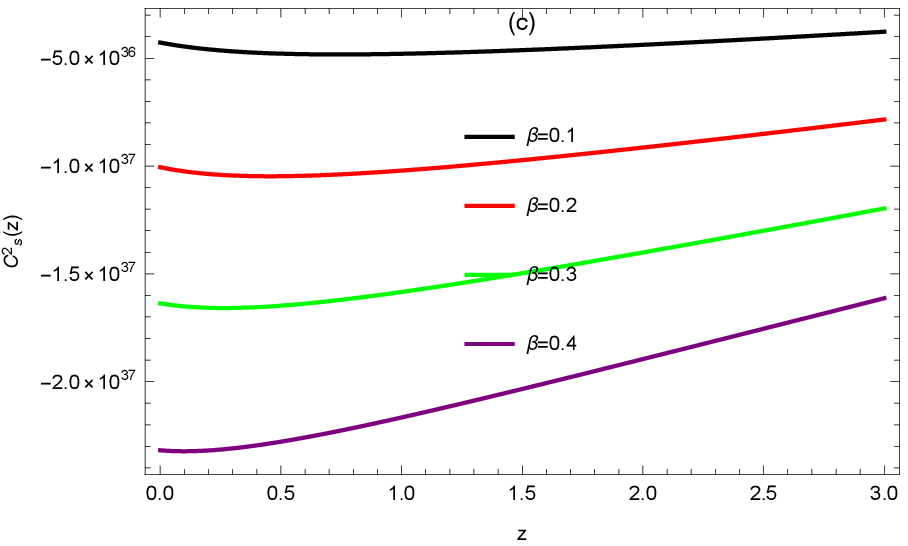,width=0.47\linewidth}\epsfig{file=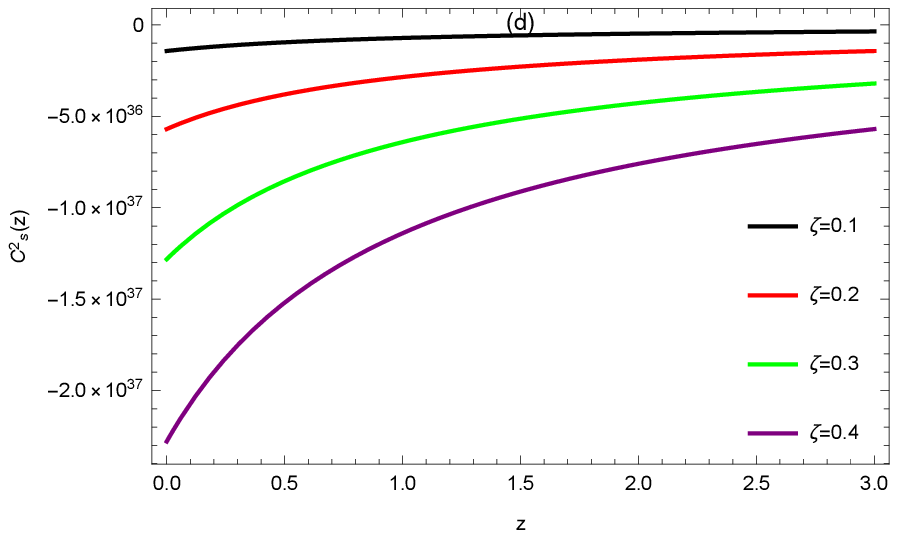,width=0.47\linewidth}
	\caption{The left (c) shows the stability analysis of model C for different values of $\beta$ while $\gamma=2$, $n=1$ and $a_{0}=0.1$ whereas the stability analysis of model D for different values of bouncing parameter when $\gamma=2$, $n=1$ and $a_{0}=0.1$ is presented in plot (d).}\label{fig14}
\end{figure}
\section{observational constraints of bouncing models}
In this section, we want to find the best fit values of bouncing model parameters ($a_{0}$, $\lambda$, $n$). For this purpose, we constrain the parameters with observational data sets. The values of parameters have been estimated by using the least square method \cite{92}. The calculated values of bouncing parameter for model B, model C and model D have been displayed in table III-V respectively. Figure 23-25 show error bar plots of Model B, Model C and Model D, respectively. It is not possible to plot an error bar plot for model A. Because the Hubble parameter for model A is defined as $H(z)=2\sqrt{\lambda}\sqrt{\log(\dfrac{1}{1+z})}$ and this expression $\sqrt{\log(\dfrac{1}{1+z})}$ gives real values only for negative values of redshift function $z$, But redshift values of Hubble data sets ranges $0<z<2.5$. These model curves have been compared with the observational data sets and $\Lambda$CDM model. $\Lambda$CDM model is defined as $H(z)^{2}=\omega_{m}(1+z)^{3}+\omega_{k}(1+z)^{2}+\omega_{r}(1+z)^{4}+\omega_{\Lambda}(1+z)^{0})^{1/2}$ where $\sum\omega_{i}=1$ and $\omega_{i}$ are free parameters of the model \cite{93}. At certain values of redshift $z$ Hubble parameter is measured from two methods.\\
 $\bullet$ Extraction of H(z) from differential ages of galaxies (DA method) \\
 $\bullet$ Estimations of H(z) from line of sight baryon acoustic oscillations (BAO).\\
 Hubble data sets points have been taken from previous literature \cite{94,95,96,97,98,99,100,101} as presented in Table IV.

To make this analysis quantified, we have employed reduced chi-squared method on $H(z)$ data set, it is defined as weighted summation of squared deviations and mathematically expressed as
\begin{equation}
\mathcal{X}_{HD}^{2}(p_{i})=\sum_{k=1}^{N_{H}}\dfrac{\bigg[H_{th}(p_{i},z_{k})-H_{obs}(z_{k})\bigg]^{2}}{\sigma_{H_{k}}^{2}}.
\end{equation}
where $H_{th}(p_{i},z_{k})$ and $H_{obs}(z_{k})$ shows theoretical and observed values of Hubble parameter. $p_{i}$ denotes model parameters e.g. (model B has two parameters that is  $\alpha$ and $a_{0}$, model C has three parameters $\beta$, $a_{0}$ and $n$, model D has only one parameter $\zeta$). $\sigma_{H_{k}}^{2}$ shows uncertainty in values of observed Hubble parameter and $z_{i}$ is redshift. Here $N_{H}$ is data points from Hubble data set. $HD$ stands for Hubble data sets from DA and BAO methods. Table III-V shows reduced chi-squared and model parameter values for model B, C and D.

\begin{table}[th!]
\begin{center}
	\begin{tabular}{ |c|c|c|c|c| }
		\hline
\rule[-1ex]{0pt}{2.5ex} Hubble data sets (HD) & $\mathcal{X}^{2}_{min}$ & parameters  \\
\hline
\rule[-1ex]{0pt}{2.5ex}		DA & 0.60873 & $\alpha=61.3453$, $a_{0}=-0.0122$  \\
\hline
\rule[-1ex]{0pt}{2.5ex}		BAO & 0.84247 &  $\alpha=57.4612$, $a_{0}=-0.0325$ \\
\hline
\rule[-1ex]{0pt}{2.5ex}	DA+BAO & 0.73011 & $\alpha=57.7235$, $a_{0}=-0.0312$\\
\hline

\end{tabular}
\end{center}
\caption{Summary of statistical analysis of model B.}
\end{table}

\begin{table}[th!]
\begin{center}
	\begin{tabular}{ |c|c|c|c|c| }
		\hline
\rule[-1ex]{0pt}{2.5ex} Hubble data sets (HD) & $\mathcal{X}^{2}_{min}$ & parameters  \\
\hline
\rule[-1ex]{0pt}{2.5ex}		DA & 0.639637  & $\beta=71.2983$, $a_{0}=-69.4779$,$n=0.4362$ \\
\hline
\rule[-1ex]{0pt}{2.5ex}		BAO & 0.88322  &  $\beta=52.6446$, $a_{0}=0.2918$, $n=0.4850$ \\
\hline
\rule[-1ex]{0pt}{2.5ex}	DA+BAO & 0.75192 & $\beta=53.1023$, $a_{0}=0.2923$, $n=0.4856$\\
\hline

\end{tabular}
\end{center}
\caption{Summary of statistical analysis of model C.}
\end{table}

\begin{table}[th!]
\begin{center}
	\begin{tabular}{ |c|c|c|c|c| }
		\hline
\rule[-1ex]{0pt}{2.5ex} Hubble data sets (HD) & $\mathcal{X}^{2}_{min}$ & parameters  \\
\hline
\rule[-1ex]{0pt}{2.5ex}		DA & 4.13944 & $\zeta=42.4589$   \\
\hline
\rule[-1ex]{0pt}{2.5ex}		BAO & 7.56656  &  $\zeta=39.6013$ \\
\hline
\rule[-1ex]{0pt}{2.5ex}	DA+BAO & 5.37551 & $\zeta=39.7871$\\
\hline

\end{tabular}
\end{center}
\caption{Summary of statistical analysis of model D.}
\end{table}

\begin{figure}[th!]
	\centering \epsfig{file=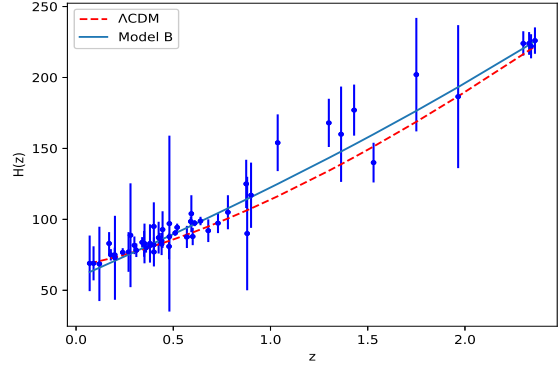, width=.47\linewidth,height=2.2in}	
	\caption{\label{Fig.4} Error bar plots of $H(z)$ datasets for matter bounce (model B).}
\end{figure}\begin{figure}[th!]
	\centering \epsfig{file=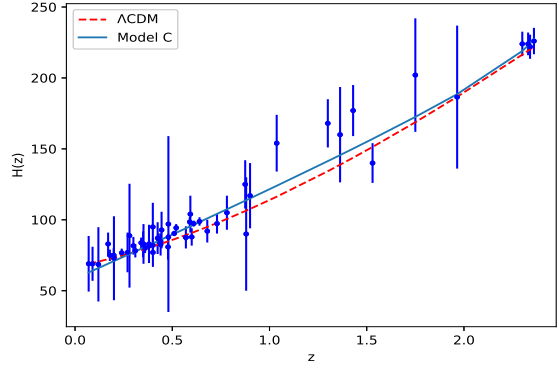, width=.47\linewidth,height=2.2in}	
	\caption{\label{Fig.4} Error bar plots of $H(z)$ datasets for power law bounce (model C).}
\end{figure}\begin{figure}[th!]
	\centering \epsfig{file=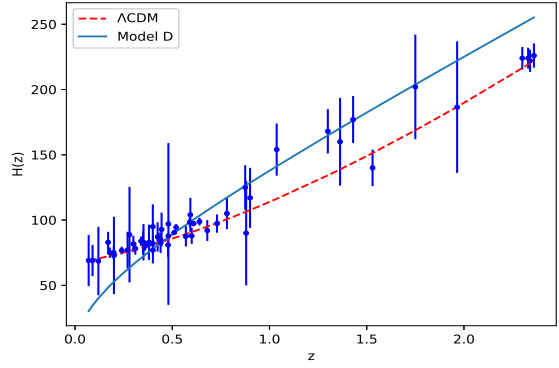, width=.47\linewidth,height=2.2in}	
	\caption{\label{Fig.4} Error bar plots of $H(z)$ datasets for oscillatory bounce (model D).}
\end{figure}

\begin{table}[th!]
\begin{center}
	\begin{tabular}{ |c|c|c|c|c| }
		\hline
\rule[-1ex]{0pt}{2.5ex} $H(z)$ & $z$ & $\sigma_{H}$ & $References$ \\
\hline
\rule[-1ex]{0pt}{2.5ex}		69 & 0.070 & 19.6 & \cite{99} \\
\hline
\rule[-1ex]{0pt}{2.5ex}		69 & 0.09 & 12 & \cite{95} \\
\hline
\rule[-1ex]{0pt}{2.5ex}	68.6 & 0.12 & 26.2 &\cite{99}\\
\hline
\rule[-1ex]{0pt}{2.5ex}	 83 & 0.17 & 8 & \cite{95}\\
\hline
\rule[-1ex]{0pt}{2.5ex}	 75 & 0.179 & 4 & \cite{97}\\
\hline
\rule[-1ex]{0pt}{2.5ex}	 75  & 0.199 & 5 & \cite{97}\\
\hline
\rule[-1ex]{0pt}{2.5ex}	 72.9 & 0.2 & 29.6 & \cite{99}\\
\hline
\rule[-1ex]{0pt}{2.5ex}	 77 & 0.27 & 14 & \cite{95}\\
\hline
\rule[-1ex]{0pt}{2.5ex}	 88.8 & 0.28 & 36.6 & \cite{99}\\
\hline
\rule[-1ex]{0pt}{2.5ex}	 83 & 0.3802 & 13.5 & \cite{101}\\
\hline
\rule[-1ex]{0pt}{2.5ex}	 83 & 0.352 & 14 &\cite{97}\\
\hline
\rule[-1ex]{0pt}{2.5ex}	 95 & 0.400  & 17 & \cite{95}\\
\hline
\rule[-1ex]{0pt}{2.5ex}	 77 & 0.4004 & 10.2 & \cite{100}\\
\hline
\rule[-1ex]{0pt}{2.5ex}	  87.1   & 0.4247 & 11.2 &\cite{96}\\
\hline
\rule[-1ex]{0pt}{2.5ex}	  92.8   & 0.44497 &  12.9 &\cite{97}\\
\hline
\rule[-1ex]{0pt}{2.5ex}	   80.9  & 0.4783 & 9 &\cite{100}\\
\hline
\rule[-1ex]{0pt}{2.5ex}	    97   &0.48 & 62 &\cite{97}\\
\hline
\rule[-1ex]{0pt}{2.5ex}	    104    & 0.593 & 13 &\cite{100}\\
\hline
\rule[-1ex]{0pt}{2.5ex}	    92   & 0.68 &  8 &\cite{97}\\
\hline
\rule[-1ex]{0pt}{2.5ex}	  105    & 0.781 & 12& \cite{97}\\
\hline
\rule[-1ex]{0pt}{2.5ex}	   125      & 0.875& 17 &\cite{96}\\
\hline
\rule[-1ex]{0pt}{2.5ex}	    90     &0.88 & 40&\cite{95}\\
\hline
\rule[-1ex]{0pt}{2.5ex}	    117  & 0.9 & 23&\cite{97}\\
\hline
\rule[-1ex]{0pt}{2.5ex}      154      &1.037& 20&\cite{95}\\
\hline
\rule[-1ex]{0pt}{2.5ex}	     168     & 1.3& 17&\cite{95}\\
\hline
\rule[-1ex]{0pt}{2.5ex}      160    &1.363 & 33.6&\cite{95}\\
\hline
\rule[-1ex]{0pt}{2.5ex}	     177     & 1.43& 18&\cite{97}\\
\hline
\rule[-1ex]{0pt}{2.5ex}	     140     & 1.53& 14&\cite{97}\\
\hline
\rule[-1ex]{0pt}{2.5ex}	     202     & 1.75& 40&\cite{97}\\
\hline
\rule[-1ex]{0pt}{2.5ex}	     186.5     & 1.965& 50.4&\cite{97}\\
\hline

\end{tabular}
\end{center}
\caption{Data Set of Hubble parameter $H(z)$ with the standard error $\pm\sigma_{H}$ from DA method.}
\end{table}

\section{Results and Summary}

The dynamics of bouncing models in 4D EGB are summerized as follows\\

$\bullet$ The four non-singular bouncing models have been investigated in the framework of 4D EGB gravity, leading us to late time cosmic acceleration.\\

$\bullet$ Evolution of scale factor, Hubble parameter, deceleration parameter, energy density, pressure and EoS parameter have been studied in detail. Kinematics of these parameters are greatly affected by bouncing parameters. The GB coupling parameter contributes less in the dynamics of EoS parameter. By choosing $\gamma=1 \times 10^{10}$ and $\gamma= 1\times 100^{120}$ or higher than these values of $\gamma$, only than we can observe the prominent contributions of 4D EGB gravity. The behaviour near bounce is primarily dependent on the bouncing parameter.\\

$\bullet$ The bouncing scale factors specify that during classical times, the cosmos is in a contracting period followed by a bounce, and then it is in an accelerating period at late times.\\

$\bullet$ All deceleration parameter values exhibit negative range and indicate accelerated expansion era, but oscillatory scale factor fails to do so, giving rise to the decelerating phase of the universe.\\

$\bullet$ The EoS parameter for all models represents phantom phase $(w<-1)$ while for  oscillatory bouncing scale factor predicts $(w>-1)$ non-phantom regime.\\

$\bullet$ Violation of Null energy conditions $(\rho+p)$ and strong energy conditions $(\rho-3p)$ near the bounce region is plotted. This is the most appropriate state for achieving non-singular bounce. Also, violation of energy condition is a clear sign that EoS parameter evolves in the phantom region $w<-1$.\\

$\bullet$ The bouncing models have been legitimized through certain cosmographic tests. The jerk, snap and lerk parameters have been found out in terms of cosmic time and redshift. The dynamics of these parameters have been presented in tabular form against cosmic time and plotted against redshift. It has been analysed through statefinder diagnostic that symmetric bounce shows $\Lambda$CDM behaviour at large value of time while statefinder values of matter bounce, super bounce and oscillatory bounce are $\bigg(0,\dfrac{2}{3}\bigg)$, $\bigg(\dfrac{3}{8},\dfrac{1}{6}\bigg)$ as $t\rightarrow\infty$ and $\bigg(0,\dfrac{1}{3}\bigg)$ as $t\rightarrow0$ respectively. It has been explored that higher derivatives of $H(z)$ represents accelerating cosmos and indulge in late time cosmic acceleration. In the case of redshift, statefinder pairs are $(1,1)$ and $(3,-15)$ for model A and model B at a larger value of time, respectively while for model C and model D $(j,s)$ $\rightarrow0$.\\

$\bullet$ Stability of bouncing models have been checked by applying the squared speed of the sound method. It has been observed that the most stable model is the matter bounce model.\\

$\bullet$ To find best-fit values, bouncing models have been constrained with DA and BAO Hubble data sets. We have calculated the values of parameters by applying the least-square fitting method \cite{92}.To make this analysis quantified, we have employed reduced chi-squared method on $H(z)$ data sets for model B, C, D and statistical results have been presentated in Table III-V respectively. Also, these curves have been plotted with observational data sets and $\Lambda$CDM model for comparison.\\

\section{Data Availability Statement}

No Data associated in the manuscript.

\section{Conflict of Interest Statement}

The authors have no competing interests to declare that are relevant to the content of this article.


\begin{thebibliography}{36}

\bibitem{1} C. M. Will, Liv. Rev. Rel. \textbf{9}, 3 (2006),


\bibitem{2} S. W. Hawking, G. F. R. Ellis, The large scale struc- ture of space time (Cambridge University Press, Cambridge, 1973).

\bibitem{3} R. Penrose, Riv. Nuovo Cimento Gravitational Collapse: The Role of General Relativity \textbf{1}, 252 (1969).

\bibitem{4} D. J. Gross, E. Witten, Nucl. Phys. B  \textbf{277}, 1 (1986).

\bibitem{5} M. C. Bento, O Bertolami, Phys. Lett. B \textbf{368}, 198 (1996).

\bibitem{6} B. Zwiebach, Phys. Lett. B \textbf{156}, 315 (1985).

\bibitem{7} D. Garfinkle, G. T. Horowitz, A. Strominger, Phys. Rev. D \textbf{43}, 3140 (1991).

\bibitem{8} G. W. Gibbons, K. Maeda, Nucl. Phys. B \textbf{298}, 741
(1988).
\bibitem{9} D. G. Boulware, S. Deser, Phys. Rev. Lett. \textbf{55}, 2656 (1985).

\bibitem{10} D. Lovelock, J. Math. Phys. \textbf{12}, 498 (1971).

\bibitem{11*} S. Nojiri, S. D. Odintsov, Phys. Let. B \textbf{631}, 1 (2005).

\bibitem{12*} S. Nojiri, S. D. Odintsov, O. Gorbunova,  J. Phys. A \textbf{39},  6627 (2006).

\bibitem{13*} T. Chiba, J. Cosmo. Astropart. Phys. \textbf{3},  107 (2005).

\bibitem{14*} C. Aïnamon, M. J. S. Houndjo, A. A. L. Ayivi, M. G. Ganiou, A. Kanfon, J. Mod. Phys. \textbf{12} 6, (2021).

\bibitem{15*} N. M. Garcia, T. Harko, F. S. N. Lobo, J. P. Mimoso, Phy. Rev. D \textbf{83}, 104032 (2011).

\bibitem{16*} M. Sharif and A. Ikram, Eur. Phys. J. C \textbf{76}, 640 (2016).

\bibitem{17*} E. Elizalde, R. Myrzakulov, V.V. Obukhov, D. Saez-Gomez, Class. Quant, Grav, \textbf{27}, 095007 (2010),

\bibitem{18*} K. Bamba, S. D. Odintsov, L. Sebastiani, S. Zerbini, Eur. Phys. J. C \textbf{67}, 295 (2010).

\bibitem{19*} A. de la Cruz-Dombriz, D. Saez-Gomez, Class. Quant. Gravi. \textbf{29}, 245014 (2012).



\bibitem{20} D. Glavan, C. Lin, 
Phys. Rev. Lett. \textbf{124}, 081301 (2020).

\bibitem{21} R. P. Woodard, Scholarpedia \textbf{10}, 32243 (2015).

\bibitem{22} S. G. Ghosh, R. Kumar, Class. Quant. Grav. \textbf{37}, 245008 (2020).

\bibitem{23} R. A. Konoplya, A. Zhidenko, Phys. Dark Univ. \textbf{30} (2020).

\bibitem{24} S. W. Wei and Y. X. Liu, Phys. Rev. D \textbf{101}, 104018 (2020).

\bibitem{25} K. Yang, B. M. Gu, S. W. Weiand Y. X. Liu, Eur. Phys. J. C \textbf{80}, 662 (2020).

\bibitem{26} S. G. Ghosh, S. D. Maharaj, Phys. Dark Univ. \textbf{30}, 100687 (2020).

\bibitem{27} A. Abdujabbarov, J. Rayimbaev, B. Turimov, F. Atamurotov, Phys. Dark Univ. \textbf{30}, 100715 (2020).

\bibitem{28} K. Jafarzade, M. Kord Zangeneh and F. S. N. Lobo, J. Cosmol. Astropart. Phys. \textbf{04}, 008 (2021).


\bibitem{29} P. G. S. Fernandes, Phys. Lett. B \textbf{805}, 135468 (2020).

\bibitem{30} S. U. Islam, R. Kumar and S. G. Ghosh, J. Cosmol. Astropart. Phys. \textbf{2009}, 030 (2020).

\bibitem{31} M. S. Churilova, Phys. Dark Univ. \textbf{31}, 100748 (2021).

\bibitem{32} X. X. Zeng, H. Q. Zhang and H. Zhang, Eur. Phys. J. C \textbf{80},
872 (2020).


\bibitem{34} K. Jusufi, A. Banerjee and S. G. Ghosh, Eur. Phys. J. C \textbf{80},
698 (2020).


\bibitem{36} A. Banerjee, T. Tangphati, P. Channuie, Astrophys. J., \textbf{909} 13, (2021).

\bibitem{37} J. M. Z. Pretel, A. Pradhan, A. Banerjee, preprint [arXiv:2108.07454].


\bibitem{39} S. Shahidi, N. Khosravi, preprint  [arXiv:2105.02372].








\bibitem {40} M. Novello, S.E.P. Bergliaffa, 
Phys. Rept. \textbf{463}, 127 (2008).

\bibitem {41} R. H. Brandenberger, 
[arXiv:1206.4196]



\bibitem {44} V. Mukhanov, Physical foundations of cosmology (Cambridge University Press, Oxford, 2005).

\bibitem {45} K. Bamba, S. D. Odintsov, 
Symmetry \textbf{7},  220 (2015).

\bibitem {46} K. Bamba, G. G. L. Nashed, W. El Hanafy, S.K. Ibraheem, 
 Phys. Rev. D \textbf {94}, 8 (2016).

\bibitem {47} W. El Hanafy and Emmanuel N. Saridakis J. Cosmol. Astropart. Phys. \textbf{09}, 019 (2021).

\bibitem {48} M. Hohmann, L. Jarv, U. Ualikhanova, 
 Phys. Rev. D \textbf{96}, 4 (2017).

\bibitem {49} J. Haro, J. Amoros, 
 J. Cosmol. Astropart. Phys.  \textbf{1412}, 12 (2014).

\bibitem {50} J . Haro, 
 J. Cosmol. Astropart. Phys., \textbf{1311}, 068 (2013).

\bibitem {51} G. Kofinas, E.N. Saridakis, 
 Phys. Rev. D \textbf{90}, 084044 (2014).

\bibitem {52}  G. Kofinas, G. Leon, E.N. Saridakis, 
 Class. Quantum Gravity \textbf{31}, 175011 (2014).

\bibitem {53} A. de la Cruz-Dombriz, G. Farrugia, J.L. Said, D.S.-C. Gomez, 
 Phys. Rev. D \textbf{97}, 10 (2018).

\bibitem {54} A. de la Cruz-Dombriz, G. Farrugia, J.L. Said, D.S.-C. Gomez,
 Class. Quantum Gravity \textbf{34}, 23 (2017).

\bibitem {55} G. Kofinas, E. N. Saridakis, 
 Phys. Rev. D \textbf{90}, 084045 (2014).

\bibitem {56} Y. F. Cai, D. A. Easson, R. Brandenberger, 
 J. Cosmol. Astropart. Phys. \textbf{1208}, 020 (2012).

\bibitem {57} R. H. Brandenberger, 
 Proc. Sci. \textbf{001}, 2010 (2010).


\bibitem {58} D. Battefeld, P. Peter, 
 Phys. Rep. \textbf{571}, 1 (2015).

\bibitem{59} A. Ilyas, M. Zhu, Y. Zheng, Y. F. Cai, E. N. Saridakis, 	J. Cosmol. Astropart.
 Phys. \textbf{09},  002 (2020).

\bibitem {60} A. Ilyas, M. Zhu, Y. Zheng, Y.F. Cai, J. High Energy Phys. \textbf{01}, 141 (2021).

\bibitem {61} M. Zhu, A. Ilyas, Y. Zheng, Y. F. Ca, E. N. Saridakis [arXiv:2108.01339].

\bibitem {62} M. Farasat Shamir,  Phys. Dark Univ. \textbf{32}, 100794 (2021).

\bibitem {63} M. Caruana, G. Farrugi , J. Levi Said, 
Eur. Phys. J. C  \textbf{80}, 640 (2020).

\bibitem {64} S. Mandal,  N. Myrzakulov, P. K. Sahoo1, R. Myrzakulov, Eur. Phys. J. Plus \textbf{136}, 760 (2021).

\bibitem {65} H. Shabani, A. H. Ziaie,
Eur. Phys. J. C \textbf{78}, 397 (2018).

\bibitem {66} K. Bamba, A. N. Makarenko, A. N. Myagky, S. Nljiri, S.D. Odintosov, J. Cosmol. Astropart. Phys. \textbf{01}, 008 (2014).

\bibitem {67} K. Bamba, A. N. Makarenko, A. N. Myagky, S.D. Odintosov, Phys. Lett. B \textbf{732}, 349 (2014).



\bibitem {70} S. Chakraborty, Phys. Rev. D \textbf{98}, 024009 (2018).

\bibitem {71} Y. F. Cai, Sci. China, Phys., Mech. Astron. \textbf{57}, 1414 (2014).

\bibitem {72} S. D. Odintsov, V. K. Oikonomou, 
 Phys. Rev. D \textbf{90}, 124083 (2014).

\bibitem {73} S. D. Odintsov, V. K. Oikonomou, 
 Phys. Rev. D \textbf{92}, 024016 (2015).

\bibitem {74} Y. B. Li, J. Quintin, D.G. Wang, Y. F. Cai, 	J. Cosmol. Astropart. Phys. \textbf{03}, 031 (2017)





\bibitem {75} Y. F. Cai, S. H. Chen, J. B. Dent, S. Dutta, E. N. Saridakis,
 Class. Quantum Gravity \textbf{28}, 215011 (2011).

\bibitem {76} J. D. Haro, Y. F. Cai, Gen. Relativ. Gravit. \textbf{47}, 95 (2015).

\bibitem {77*} P. Sahoo, S. Bhattacharjee, S. K. Tripathy, P. K. Sahoo, Mod. Phys. Lett. A \textbf{2050095}, 14  (2020).

\bibitem {77} C. Cattoen, M. Visser,
 Class. Quantum Gravity \textbf{22}, 4913–4930 (2005).

\bibitem {78} S. Capozziello, S. Nojiri, S.D. Odintsov,  Phys. Lett. B \textbf{781}, 99-106  (2018).

\bibitem {79} U. Alam, V. Sahni, T. D. Saini, A. A. Starobinsky, 
 Mon. Not. R. Astron. Soc. \textbf{344}, 1057 (2003).

\bibitem {80} A. G. Riess, et al., Astrophys. J. \textbf{607}, 665 (2004).





\bibitem{84} V. Sahni, T. D. Saini, A. A. Starobinsky, U. Alam, JETP Lett.\textbf{77}, 201-206(2003)

\bibitem {85} F. Y. Wang, Z. G. Dai1, Shi Qi, Astron. Astrophys. \textbf{507}, 53 (2009).


\bibitem{86} D. Rapetti, S. W. Allen, M. A. Amin, R. D. Blandford, Mon. Not. Roy. Astron. Soc. \textbf{375}, 1510, (2007).

\bibitem{87} A. Mukherjee, N. Banerjee, Astrophys. Space Sci. \textbf{352}, 893 (2014).

\bibitem{88} A. Aviles, C. Gruber, O. Luongo, H. Quevedo, Phys. Rev. D \textbf{86}, 123516 (2012).

\bibitem{89} S. K. Tripathy, R. K. Khuntia, and P. Parida, Eur. Phys. J. Plus  \textbf{134}, 504 (2019).

\bibitem{90} A. S. Agrawal, L.  Pati, S.K. Tripathy, B. Mishra, Phys. Dark Universe \textbf {33}, 100863 (2021).

\bibitem{91} S. Mandal, S. Bhattacharjee, S. K. J. Pacif, P.K. Sahoo, Phys. Dark Universe, \textbf{28}, 100551 (2020).

\bibitem{92} D. S. Wilks,  Statistical Method in the Atomspheric Sciences; Elsevier Inc.: Burlington, MA, USA, 2006;.

\bibitem{93} G. S. Sharov, E. S. Sinyakov, Math. Model. Anal. \textbf{8}, 1 (2020).

\bibitem{94} N. Suzuki, et al.,  Astrophys. J. \textbf{746}, 85 (2012).

\bibitem{95} J. Simon, L. Verde, R. Jimenez, Phys. Rev. D \textbf{71}, 123001 (2005) .

\bibitem{96} D. Stern, et al., J. Cosmol. Astropart. Phys. \textbf{02}, 008 (2010).

\bibitem{97} M. Moresco, et al., J. Cosmol. Astropart. Phys. \textbf{08}, 006 (2012).

\bibitem{98} N. G. Busca, et al. [arXiv:1211.2616].

\bibitem{99} C. Zhang, et al., Res. Astron. Astrophys. \textbf{14}, 1221 (2014).

\bibitem{100} C. Blake, et al. [arXiv:1204.3674].

\bibitem{101} C. H. Chuang, Y. Wang [arXiv:1209.0210].













































\end{thebibliography}
\end{document}